\documentclass[preprint]{aastex631}

\graphicspath{{./}{figures/}}

\usepackage{graphicx}

\newcommand{\OIII}{[O\,{\tiny III}]}

\newcommand{\NII}{[N\,{\tiny II}]}

\begin{document}

\title{A {\it Chandra} X-ray Survey of Optically Selected Close Galaxy Pairs: Unexpectedly Low Occupation of Active Galactic Nuclei}

\author[0000-0002-7875-9733]{Lin He}
\affiliation{School of Astronomy and Space Science, Nanjing University, Nanjing 210023, China}
\affiliation{Key Laboratory of Modern Astronomy and Astrophysics, Nanjing University, Nanjing 210023, China}

\author[0000-0001-9062-8309]{Meicun Hou}
\affiliation{Kavli Institute for Astronomy and Astrophysics, Peking University, Beijing 100871, China}
\email{houmc@pku.edu.cn}

\author[0000-0003-0355-6437]{Zhiyuan Li}
\affiliation{School of Astronomy and Space Science, Nanjing University, Nanjing 210023, China}
\affiliation{Key Laboratory of Modern Astronomy and Astrophysics, Nanjing University, Nanjing 210023, China}

\author[0000-0002-9767-9237]{Shuai Feng}
\affiliation{College of Physics, Hebei Normal University, 20 South Erhuan Road, Shijiazhuang, 050024, China}
\affiliation{Hebei Key Laboratory of Photophysics Research and Application, 050024 Shijiazhuang, China}

\author[0000-0003-0049-5210]{Xin Liu}
\affiliation{Department of Astronomy, University of Illinois at Urbana-Champaign, Urbana, IL 61801, USA}
\affiliation{National Center for Supercomputing Applications, University of Illinois at Urbana-Champaign, 605 East Springfield Avenue, Champaign, IL 61820, USA}

\begin{abstract}
High-resolution X-ray observations offer a unique tool for probing the still elusive connection between galaxy mergers and active galactic nuclei (AGNs). We present an analysis of nuclear X-ray emission in an optically selected sample of 92 close galaxy pairs (with projected separations $\lesssim 20$ kpc and line-of-sight velocity offsets $<$ 500~km~s$^{-1}$) at low redshift ($\bar{z} \sim 0.07$), based on archival {\it Chandra} observations.
The parent sample of galaxy pairs is constructed without imposing an optical classification of nuclear activity, thus is largely free of selection effect for or against the presence of an AGN. Nor is this sample biased for or against gas-rich mergers.
An X-ray source is detected in 70 of the 184 nuclei, giving a detection rate of $38\%^{+5\%}_{-5\%}$, down to a 0.5--8 keV limiting luminosity of $\lesssim 10^{40}\rm~erg~s^{-1}$. 
The detected and undetected nuclei show no systematic difference in their host galaxy properties such as galaxy morphology, stellar mass and stellar velocity dispersion.  
When potential contamination from star formation is avoided (i.e., $L_{\rm 2-10~keV} > 10^{41}\rm~erg~s^{-1}$), the detection rate becomes $18\%^{+3\%}_{-3\%}$ (32/184), 
which shows no excess compared to the X-ray detection rate of a comparison sample of optically classified single AGNs.
The fraction of pairs containing dual AGN is only $2\%^{+2\%}_{-2\%}$.
Moreover, most nuclei at the smallest projected separations probed by our sample (a few kpc) have an unexpectedly low apparent X-ray luminosity and Eddington ratio, which cannot be solely explained by circumnuclear obscuration.
These findings suggest that close galaxy interaction is not a sufficient condition for triggering a high level of AGN activity. 
\end{abstract}


\defcitealias{Hou_2020}{H20}
\defcitealias{Liu_2011}{L11}
\defcitealias{Feng_2019}{F19}

\section{Introduction} \label{sec:intro}
It is a generic prediction of the standard paradigm of hierarchical structure formation that most galaxies frequently interact with other galaxies during their lifetime.
When the two interacting galaxies are gravitationally bound, their ultimate fate is to merge, eventually forming a more massive galaxy.
In the course of galaxy mergers, tidal force and ram pressure act to significantly redistribute the stellar and gaseous contents of the interacting pair \citep{Toomre_1972,Barnes_1992}. 
It is theoretically predicted and has been demonstrated by numerical simulations (e.g., \citealp{DiMatteo_05}) that upon close passages, gravitational torques drive gas inflows to the center of one or both galaxies, potentially triggering nuclear star formation and active galactic nuclei (AGNs). 
 A physical consequence
of this scenario is the prevalence of AGN pairs 
in (major) galaxy mergers, which involve two SMBHs with simultaneous active accretion.
Specifically, ``dual AGNs'', AGN pairs with a separation $\lesssim$ 10 kpc in projection, are generally expected at the intermediate- to late-stage of mergers (see recent review by \citealp{DeRosa_19}). This is a crucial phase during which the SMBH(s) can significantly grow its mass, preceding the formation of a SMBH binary and their ultimate merger \citep{Merritt_2005}.

As observational validation of the above scenario, a number of systematic searches for dual AGN candidates have been conducted over the past decade, primarily in the optical band thanks to wide-field, homogeneous spectroscopic surveys such as the Sloan Digital Sky Survey (SDSS). In particular, the search for galactic nuclei with double-peaked narrow emission lines (e.g., [O III]; \citealp{Wang_09,Liu_10}) aims at tight AGN pairs (typically 1--10 kpc in separation, but even less) that pertain to the late stage of merger, whereas the search for resolved pairs of galactic nuclei both showing the optical emission line characteristics of Syfert or LINER (Low Ionization Nuclear Emission-line Region) covers larger projected separations up to $\sim$100 kpc \citep[][hereafter L11]{Liu_2011}. Confirmation of the AGN nature in these optically-selected candidates, however, often require follow-up observations in the X-ray and/or radio bands \citep{Comerford,Silverman2011,Teng_12,Liu_13,Fu2015a,Fu2015,Brightman_18,Gross2019,Hou_19,Foord2020}, which are generally thought to trace immediate radiation from the SMBH (more precisely, from the accretion disk, corona and/or jets) and tend to be more immune to circumnuclear obscuration. 
Infrared observations have also played an effective role in revealing dual AGNs, especially in gas-rich merging systems, which tend to select highly obscured AGNs \citep{Satyapal_14,Satyapal_17,Pfeifle_19}.
An alternative approach \citep{Koss_12} starts with a hard X-ray ($\gtrsim 10$ keV) AGN detected in the Swift/BAT survey and tries to associate it with another AGN in a companion galaxy within a projected distance of 100 kpc, if existed. 
This approach, however, inevitably introduced selection bias toward X-ray-luminous AGNs due to the moderate sensitivity of Swift/BAT.  
Nevertheless, these approaches have achieved a certain degree of success, revealing a growing number of AGN pairs and dual AGNs. 

Clearly, having a sizable and unbiased sample of genuine dual AGNs is crucial for a thorough understanding of the causality between galaxy mergers and AGN triggering.
Recently, \citet[][hereafter H20]{Hou_2020} carried out a systematic search for X-ray-emitting AGN pairs, using archival {\it Chandra} observations and based on the \citetalias{Liu_2011} sample of $\sim 10^3$ optically-selected AGN pairs at low redshift (with a median redshift $\bar{z} \sim 0.1)$. 
Thanks to the superb angular resolution of {\it Chandra}, unattainable from any other X-ray facility, one can unambiguously resolve and localize the putative AGN even in close pairs.
More importantly, the typical sensitivity of {\it Chandra} observations used by \citetalias{Hou_2020} is sufficient to probe low-luminosity AGNs (i.e., weakly accreting SMBHs) down to a limiting 2--10 keV X-ray luminosity of $L_{2-10} \sim 10^{40}{\rm~erg~s^{-1}}$, which is necessary for a complete census of nuclear activity.

Among 67 pairs of the optically-selected AGN candidates with useful {\it Chandra} data, \citetalias{Hou_2020} found that 21 pairs show significant X-ray emission from both nuclei (i.e., probable AGN pairs), with an additional 36 pairs having only one of the two nuclei detected. 
The X-ray detection rate of all 134 nuclei, $58\% \pm 7\%$ (1$\sigma$ Poisson errors), is significantly higher than that ($17\% \pm 4\%$) of a comparison sample of star-forming galaxy pairs, classified also based on optical emission line ratios.
Moreover, interesting trends were revealed for the mean X-ray luminosity as a function of the projected separation, $r_{\rm p}$, which is taken as a proxy for the merger phase, where larger (smaller) $r_{\rm p}$ represents the earlier (later) stage of a merger. 
First, $L_{2-10}$ increases with decreasing projected separation in AGN pairs at $r_{\rm p} \gtrsim 20$ kpc, suggesting enhanced SMBH accretion even in early-stage mergers, perhaps related to the first pericentric passage of the two galaxies.
Secondly and unexpectedly, $L_{2-10}$ decreases (rather than increases) with decreasing $r_{\rm p}$ at $r_{\rm p} \lesssim 10$ kpc, which appears contradicting with the intuitive expectation that tidal force-driven gas inflows become more and more prevalent as mergers proceed. 
Despite the small number statistics, \citetalias{Hou_2020} proposed two physical explanations for this latter behavior: (i) merger-induced gas inflows become so strong that an enhanced central concentration of cold gas heavily obscures even the hard (2--10 keV) X-rays; (ii) AGN feedback triggered by the first pericentric passage acts to expel gas from the nuclear region and consequently suppress or even halt SMBH accretion.
The latter possibility is of particular interest, potentially offering insight into the still elusive processes of SMBH feeding and feedback during an indispensable stage of galaxy evolution.

Extending the study of \citetalias{Hou_2020}, in this work we use archival {\it Chandra} observations to survey the nuclear X-ray emission from a new sample of close galaxies pairs.
These close galaxies pairs are selected from optical spectroscopic surveys (see Section \ref{sec:sample} for details), but they are not subject to a pre-selection of optical AGN characteristics as applied in \citetalias{Hou_2020}, thus allowing for an unbiased view of AGN activity through their nuclear X-ray emission.  
This paper is structured as follows. 
Section~\ref{sec:sample} describes the construction of a new sample of close galaxy pairs with archival {\it Chandra} observations. 
Data analysis toward detection and characterization of the nuclei are detailed in Section~\ref{sec:analysis}.
Section~\ref{sec:result} presents the results, including the properties and statistics of the X-ray detected nuclei and a reexamination of the behavior of $L_{\rm 2-10}$ as a function of $r_{\rm p}$.
Section~\ref{sec:discussion} summarizes the study and address most significant implications. 
Throughout this work, weassume a concordance cosmology with $\Omega_m = 0.3$, $\Omega_{\Lambda} = 0.7$, and $H_{0}=70$ km s$^{-1}$ Mpc$^{-1}$.
Errors are quoted at 1$\sigma$ confidence level unless otherwise stated.

\section{The Sample of Close Galaxy Pairs}
\label{sec:sample}

In this work, we construct a new sample of close galaxy pairs based on the parent sample of galaxy pairs recently presented by \citet[][hereafter F19]{Feng_2019}.
The \citetalias{Feng_2019} sample itself was extracted from the SDSS DR7 \citep{Abazajian2009} photometric galaxy catalog, with $\sim$95\% of the cataloged galaxies having an available spectroscopic redshift, which was primarily from SDSS and supplemented by LAMOST \citep{Luo2015,Shen2016}, GAMA \citep{Baldry2018} and other spectroscopic surveys (see detailed description in \citetalias{Feng_2019}). 
A close galaxy pair was selected if the two member galaxies have a line-of-sight velocity offset $\Delta v < 500\rm~km~s^{-1}$ and a projected separation $r_{\rm p} \lesssim 20$ kpc.
We also required that each galaxy has only one neighbor galaxy with a similar redshift within a projected separation of 100 kpc and a velocity offset of 500 km s$^{-1}$, to minimize environmental effects typical of compact groups or clusters. 
Contrary to \citetalias{Feng_2019}, who focused on pairs with $r_{\rm p} > 10h^{-1}$ kpc, we impose no lower limit on $r_{\rm p}$. However, due to the resolution limit of the optical surveys ($\sim 1 \arcsec$), the \citetalias{Feng_2019} sample still suffers from incompleteness for the most closely separated pairs (i.e., $\lesssim$ 1 kpc).

We thus have a preliminary list of 3337 close galaxy pairs. 
A comparison with the \citetalias{Liu_2011} sample of optically-selected AGN pairs shows that the two samples have 130 common pairs, whereas 3207 pairs are in the \citetalias{Feng_2019} sample but not in the \citetalias{Liu_2011} sample. 
This difference partly stems from the fact that the \citetalias{Liu_2011} sample, which was primarily based on SDSS DR7 spectroscopic redshifts, suffers from the restriction of SDSS fiber collision and thus is missing closely separated galaxies pairs. 
The \citetalias{Feng_2019} sample was exactly designed to overcome this incompleteness, thereby significantly increasing the number of close galaxy pairs. 
Moreover, the \citetalias{Liu_2011} sample required both galaxies in a pair to have a Seyfert or LINER classification based on the optical emission line diagnostics, whereas the \citetalias{Feng_2019}  sample only required a spectroscopic redshift based primarily on stellar continuum, thus in principle minimizes the selection bias for or against AGN activity in closely interacting galaxies (though see Section \ref{subsec:lxrp} for potential bias for the most luminous AGNs in a few {\it Chandra} observations), as well as selection bias for or against gas-rich mergers.

We cross-matched the \citetalias{Feng_2019} sample with the {\it Chandra} X-ray data archive to select pairs with observations taken with the Advanced CCD Imaging Spectrometer (ACIS) and publicly available as of June 2022. 
Similar to \citetalias{Hou_2020}, we requested that both galactic nuclei in a pair fall within the ACIS field-of-view (FoV) and within $8\arcmin$ from the aimpoint, to ensure the feasibility of source detection and photometry. 
We further visually inspected the SDSS to filter several spurious galaxy pairs, which are most likely compact star-forming clusters/complexes that mimicked a second galactic nucleus.
Our final sample consists of 92 optically and X-ray selected galaxy pairs, which have $r_{\rm p}$ ranging from 3.0 kpc to 19.7 kpc.
This small fraction (92/3337) reflects the empirical rule that on average only a few percent of randomly selected sky targets would fall on a {\it Chandra}/ACIS footprint.
Basic information of these galaxy pairs are given in Table \ref{tab:info}. 


Our sample is an extension of the AGN pairs and SFG pairs studied by \citetalias{Hou_2020}. 
The \citetalias{Hou_2020} AGN pairs, selected from the parent sample of \citetalias{Liu_2011}, cover a wider range of projected separations ($r_{\rm p} < 100 $ kpc) and have both nuclei classified as an AGN based on the optical emission line diagnostics. 
\citetalias{Hou_2020} also constructed a comparison sample of SFG pairs (i.e., both nuclei having the optical emission line diagnostics of star formation). 
Considering only the close pairs (i.e., those with $r_{\rm p} < 20 $ kpc) in \citetalias{Hou_2020}, there are 28 AGN pairs and 12 SFG pairs. 
For clarity, hereafter we refer to AGN pairs or SFG pairs of \citetalias{Hou_2020} as those pairs with $r_{\rm p} < 20$ kpc only, unless otherwise stated.
With our new sample, which presumes no distinction between optically classified AGN and SFG, the total number of close galaxy pairs with both {\it Chandra} and optical spectroscopic observations is now more than doubled. 
We note that the new sample includes 17 AGN pairs and 1 SFG pair in \citetalias{Hou_2020}. These pairs are kept in the following analysis, but caution is taken not to double-count them when an analysis also involves those pairs from \citetalias{Hou_2020}.
These also exist some pairs which belong to \citetalias{Hou_2020} but are not included in the new sample. This is mainly due partly to the fact that the \citetalias{Liu_2011} sample 
did not impose the requirement on the absence of a third galaxy within 100 kpc and also included some pairs that are not part of the parent galaxy sample of \citetalias{Feng_2019}.

\begin{figure}
\centering
\includegraphics[width=0.95\textwidth]{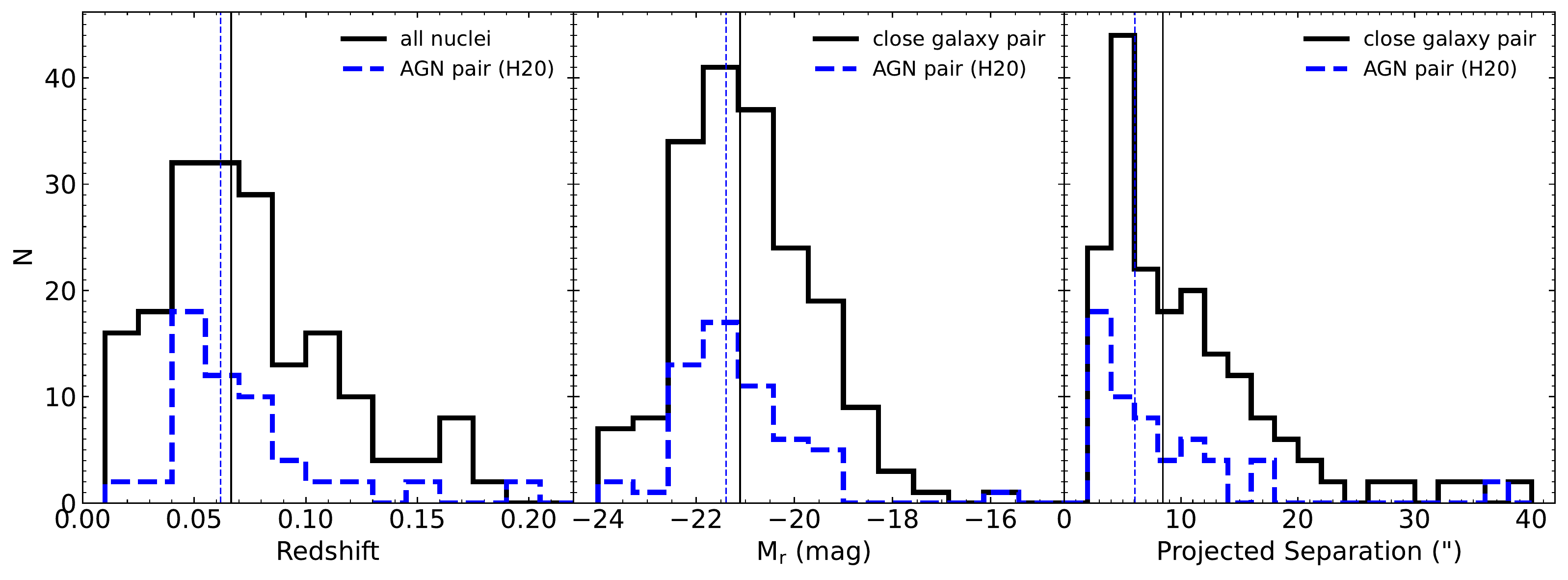}
\caption{Redshift ({\it left panel}), $r$-band absolute magnitude ({\it middle panel}) and projected angular separation ({\it right panel}) distributions of the close galaxy pairs studied in this work (black solid histogram), in comparison with the close AGN pairs (blue dashed) in \citetalias{Hou_2020}. The vertical lines mark the median value of the individual samples.
}
\label{fig:samples}
\end{figure}

Figure \ref{fig:samples} compares the redshift (left panel) and SDSS $r$-band absolute magnitude ($M_r$; middle panel) distributions of the current sample with those of the AGN pairs in \citetalias{Hou_2020}. 
The current sample has a median redshift $\bar{z} = 0.067$ and a median $r$-band absolute magnitude $\bar{M}_r = -21.1 \rm~mag$, while the close AGN pairs have a similar $\bar{z} = 0.062$ and $\bar{M}_r = -21.4 \rm~mag$.



\section{Data Analysis}\label{sec:analysis}

\subsection{{\it Chandra} Data Preparation}\label{subsec:X-ray data}

The {\it Chandra}/ACIS data were reprocessed following the standard procedure, using CIAO v4.13 with the calibration files CALDB v4.9.5\footnote{\url{http://cxc.harvard.edu/ciao/}}. 
Among the 92 galaxy pairs in the current sample, 78 pairs have only one observation, while the the other 14 pairs have been observed more than one time, for which we combined all available observations. 

Following the procedures in \citetalias{Hou_2020}, for each observation we produced counts, exposure and point-spread function (PSF) maps on the natal pixel scale of $0 \farcs 492$ in the 0.5--2 ($S$), 2--8 ($H$) and 0.5--8 ($F$) keV band. 
The exposure maps and the PSF maps were weighted by a fiducial incident spectrum, which is an absorbed power-law with a photon-index of 1.7 (a median value for AGN, see \citealp{Winter2009}) and absorption column densities $N_{\rm H} = 10^{22}\rm~cm^{-2}$ for the $H$ band and $N_{\rm H} = 10^{21}\rm~cm^{-2}$ for the $S$ band.

For targets with multiple observations, the counts, exposure and PSF maps of individual observations were reprojected to a common tangential point after calibrating their relative astrometry, to produce combined images that maximize the source detection sensitivity. 
Only the I0, I1, I2, I3 CCDs for the ACIS-I observations and the S2, S3 CCDs for the ACIS-S observations were included at this step.
We have examined the light curves of each observation and filtered time intervals contaminated by significant particle flares, if any. 
The effective exposure time of each target pair ranged from 1.1 to 240.1 ks, with a median value of 13.7 ks.

\subsection{X-ray Counterparts and Photometry}\label{subsec:detection}
We followed the procedures detailed in \citetalias{Hou_2020} to search for X-ray counterparts of the optical nuclei in our close galaxy pairs.
We first performed source detection in the 0.5--2, 2--8 and 0.5--8 keV bands for each galaxy pair using the CIAO tool {\it wavdetect}, with the 50\% enclosed count fraction (ECF) PSF maps supplied and a false detection probability of $10^{-6}$. 
We then searched for an X-ray counterpart of each optical nucleus from the X-ray source lists output by {\it wavdetect}, adopting a matching radius of 2\arcsec,
an empirically optimal value given the angular resolution and astrometry.
This is further justified by a random matching test by artificially shifting the positions of all nuclei by $\pm 10\arcsec$ in R.A. and Decl., which finds on average less than one coincident match with the detected X-ray sources.
We note that no pair in our sample has angular separation less than this matching radius (see the third panel in Figure \ref{fig:samples}), which means two nuclei in a pair would not be matched with one identical X-ray counterpart in any case.
If the optical nucleus was matched with an X-ray counterpart in any of the three energy bands, we consider it as {\it X-ray-detected}. 

Source photometry was then calculated using the CIAO tool {\it aprate}, which properly handles the counting statistics in the low-count regime.
Source count at a given band was extracted from within the 90\% enclosed count radius (ECR).
The local background was evaluated from a concentric annulus with inner-to-outer radii 2--5 times the 90\% ECR for the inner radius, excluding pixels falling within the 90\% ECR of neighboring sources, if any.
In a few cases where the two nuclei have overlapping 90\% ECR, we adopt the 50\% ECR for photometry.
The net photon flux was derived by dividing the exposure map and corrected for the ECF. 

For the optical nuclei without an X-ray counterpart found by {\it wavdetect}, we extracted the source and background counts in a similar way and estimated a 3$\sigma$ upper limit of the net photon flux using {\it aprate}.
If the 3$\sigma$ lower limit were greater than zero, the nucleus is regarded as an X-ray detection. Using this more quantitative criterion, we recover a few more nuclei with significant X-ray emission that have been filtered by {\it wavdetect}. 
For the remaining nuclei, we again used {\it aprate} to derive a 3$\sigma$ upper limit of the net photon flux.

\begin{figure}[htbp]\centering
  \includegraphics[width=0.7\linewidth]{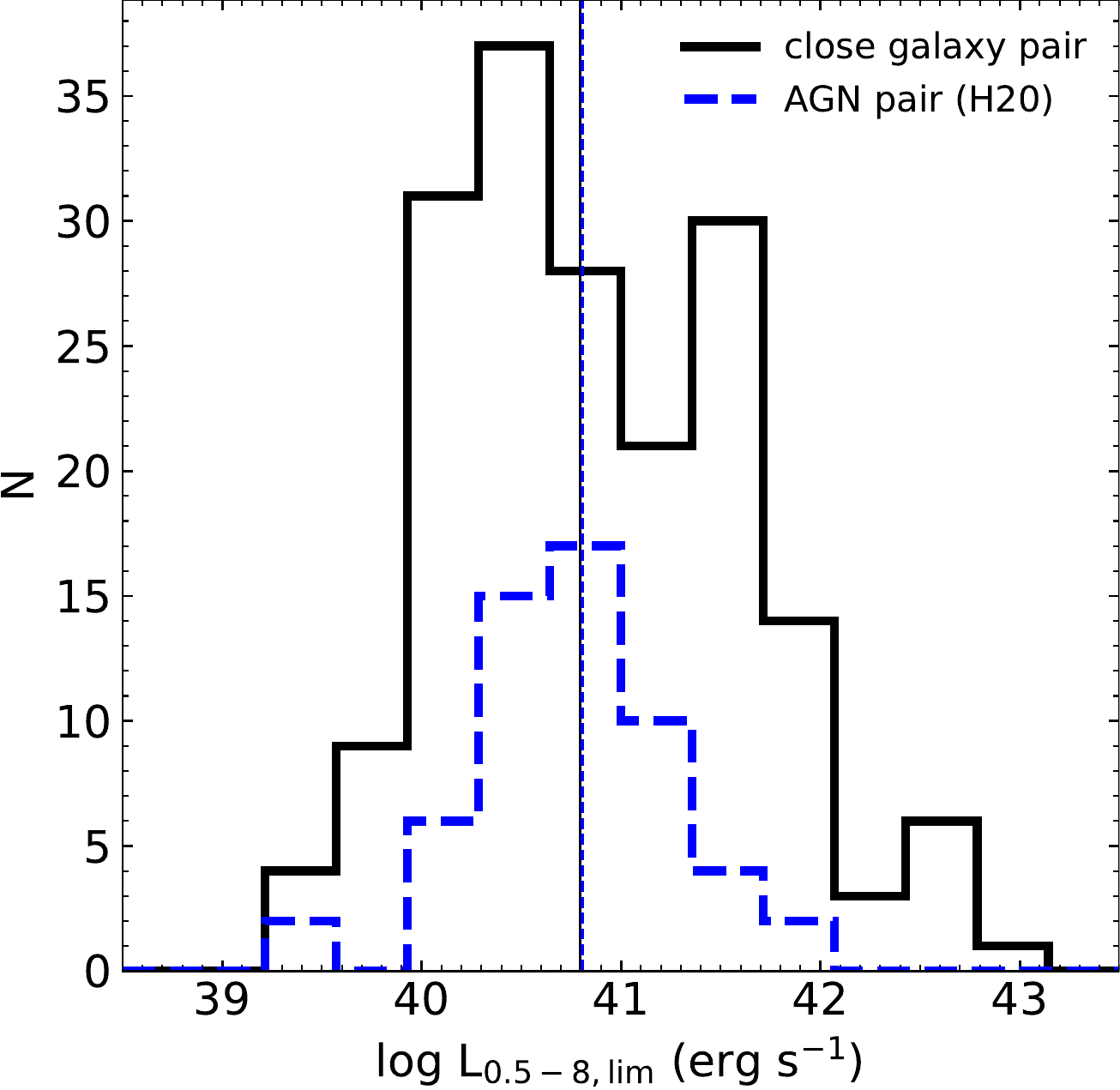}  
\caption{0.5--8 keV detection limit distribution of the close galaxy pairs studied in this work (black solid histogram), in comparison with the close AGN pairs (blue dashed) in \citetalias{Hou_2020}. The vertical lines mark the median value of the individual samples.
}
\label{fig:Llim}
\end{figure}

The net photon fluxes (or upper limits) were then converted to an unabsorbed luminosity in the 0.5--2 and 2--10 keV bands, by multiplying a unique conversion factor for a given energy band according to the fiducial incident spectrum described in Section~\ref{subsec:X-ray data}. 
The net counts, photon fluxes and luminosities are listed in Table \ref{tab:Xray}.
We have also determined the detection limit of a given band at the position of each nucleus, following the method of \citet{Kashyap_10}.
Figure~\ref{fig:Llim} plots the histogram of the 0.5--8 limiting luminosity for both the current sample (listed in Table \ref{tab:info}) and the AGN pairs of \citetalias{Hou_2020}, which have a similar distribution, facilitating a direct comparison between the two sample.

\subsection{{\it NuSTAR} spectral analysis}\label{subsec:nustar}
To help constrain the presence of intrinsically luminous but heavily obscured AGNs in the sample galaxies, we utilized archival {\it NuSTAR} observations that are sensitive to the hard ($\gtrsim$ 10 keV) X-rays from obscured AGNs. 
Eight pairs in the current sample have been observed by {\it NuSTAR}, with an effective exposure ranging from 19.5 ks to 211.3 ks.
We note that half of these eight observations were taken as a targeted observation to probe the hard X-ray emission from a putative AGN.

The {\it NuSTAR} data were downloaded and reprocessed following the standard {\it nupipeline} in the software package NuSTARDAS v2.1.2. The spectra of each galaxy pair were extracted for both focal plane modules A and B (FPMA and FPMB) with {\it nuproducts}. 
A circular region was used to extract the source spectrum, which has a radius of 60$\arcsec$, approximately equalling to 75\% ECR.
Since the two nuclei in a given pair are not well resolved by {\it NuSTAR}, the source center was set to be the brighter nucleus as seen by {\it Chandra}, which is generally consistent with the peak of the {\it NuSTAR}-detected signal.
The background spectra were extracted from a concentric annulus with an inner radius of 90$\arcsec$ and an outer radius of 150$\arcsec$. 
It turns out that three of the eight pairs show no significant signal above the background, thus they were neglected in the spectral analysis.

For the five pairs with significant hard X-ray emission, the spectra were grouped to achieve a signal-to-noise ratio (S/N) greater than 3 per bin over the energy range of 3--79 keV. 
We follow the method of \citep{Zappacosta2018} to simulate background spectrum using the software NUSKYBGD \citep{Wik2014} to account for the spatially dependent background of NuSTAR. This task aims to compute the relative strengths of different background components and hence well reproduce the background spectrum at each position of the detector. 
The FPMA/FPMB spectra were jointly fitted.
Spectral analysis was carried out with Xspec v.12.12.1c, adopting the $\chi^2$ statistics to determine the best-fit model.
Since we are mainly interested in constraining the line-of-sight absorption column density and the intrinsic X-ray luminosity, 
we adopted a phenomenological model, an absorbed power-law {\it tbabs*powerlaw}, as the default model.
In one source, J1451+0507, significant excess is present around 6.4 keV, which can be interpreted as an iron fluorescent emission line often seen in luminous AGNs. For this source we added a Gaussian component to account for the putative Fe line, which significantly improved the fit.
Such a Gaussian component was not required for the other four sources. 
The spectral fit results are listed in Table \ref{tab:Nuspec}, which include the best-fit absorption column density, photon-index, 3--79 keV unabsorbed flux and 2--10 keV unabsorbed luminosity converted from the best-fit model.

\section{Results}\label{sec:result}

\subsection{X-ray Detection Rate}\label{subsec:detectionrate}

A bright X-ray source matched with the galactic nucleus usually refers to an AGN, but with potential contamination from the host galaxy (i.e., nuclear starburst). As estimated in Section~\ref{subsec:properties}, The X-ray emission due to star formation is neglectable compared to AGN, especially for the nuclei with $L_{\rm 2-10} > 10^{41} \rm~erg~s^{-1}$.

\begin{figure}[htbp]\centering
\includegraphics[width=1.0\textwidth]{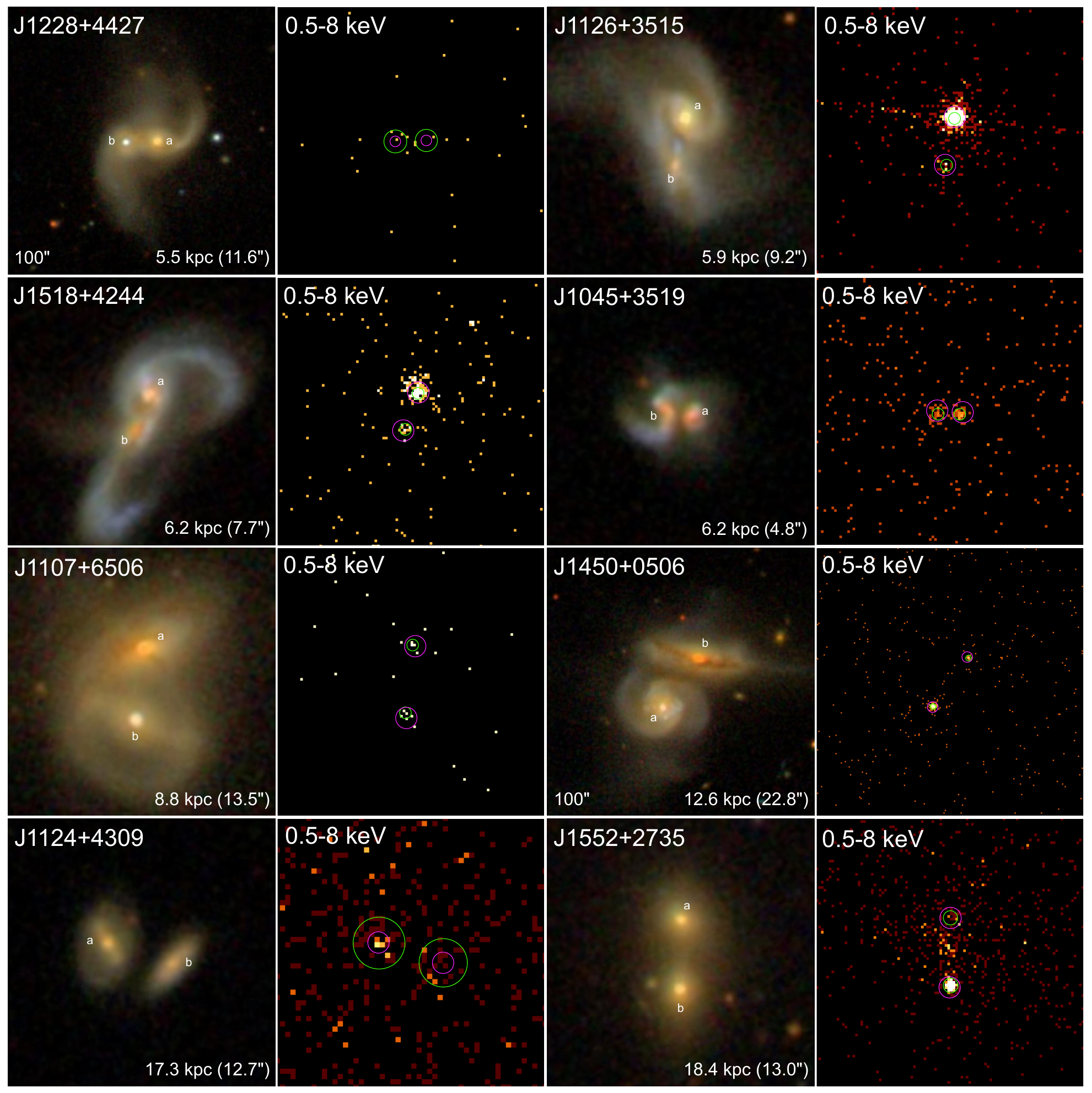}
\caption{SDSS $gri$-color composite (first and third columns) and {\it Chandra}/ACIS 0.5--8 keV (second and fourth columns) images of the 8 newly-found close galaxy pairs with both nuclei detected in the X-rays. Each panel has a size of $50\arcsec \times 50\arcsec$ unless otherwise labeled. North is up and east is to the left. Magenta circles denote positions of the optical nuclei. Green circles represent the 90\% ECR of local PSF.
}
\label{fig:image_closepair}
\end{figure}

In total, we find 70 X-ray-detected nuclei, among which 67 are detected in the $S$ band, 58 are detected in the $H$ band, and 70 are detected in the $F$ band.
Among the 92 close galaxy pairs, 14 pairs have both nuclei detected, 42 pairs have only one of the two nuclei detected, and 36 pairs have no X-ray detection. 
Figure \ref{fig:image_closepair} displays the SDSS $gri$ color-composite images and the {\it Chandra} 0.5--8 keV images of the 8 newly-found close galaxy pairs with both nuclei detected in the X-rays (the other 6 pairs have been studied and presented in \citetalias{Hou_2020}).
These 16 nuclei have a 0.5--8 keV  luminosity ranging from $1.3\times10^{40}- 6.3\times10^{42}\rm~erg~s^{-1}$.

We find an X-ray detection rate of $38\%^{+5\%}_{-5\%}$ (70/184) among the 92 close galaxy pairs.
The quoted error takes into account the counting (Poisson) error in both the numerator and denominator.
In the more conservative case where we only consider X-ray counterparts with a 2--10 keV unabsorbed luminosity $L_{\rm 2-10} > 10^{41} \rm~erg~s^{-1}$, which are most likely dominated by an AGN (see Section~\ref{subsec:properties}), the detection rate becomes $18\%^{+3\%}_{-3\%}$ (32/184).
For comparison, \citetalias{Hou_2020} gave a detection rate of $27\%^{+5\%}_{-5\%}$ (36/134) for their entire sample of AGNs (i.e., regardless of the value of $r_{\rm p}$) above the threshold of $L_{\rm 2-10} = 10^{41} \rm~erg~s^{-1}$.
This factor of $\sim 1.5$ difference may be understood as a systematically higher fraction of true AGNs in the \citetalias{Hou_2020} sample, which is consistent with their original optical classification.
When considering the fraction of pairs containing at least one X-ray-detected nucleus with $L_{\rm 2-10} > 10^{41}$, we find $32\%^{+7\%}_{-7\%}$ (30 out of 92 pairs) for the current sample, which is again somewhat lower than that of the \citetalias{Hou_2020} AGN sample ($47\%^{+11\%}_{-10\%}$; 32 out of 67 pairs).
These and additional detection rates are reported in Table~\ref{tab:rate}.

\subsection{Global X-ray Properties}
\label{subsec:properties}

\begin{figure}[htbp]\centering
\includegraphics[width=0.48\linewidth]{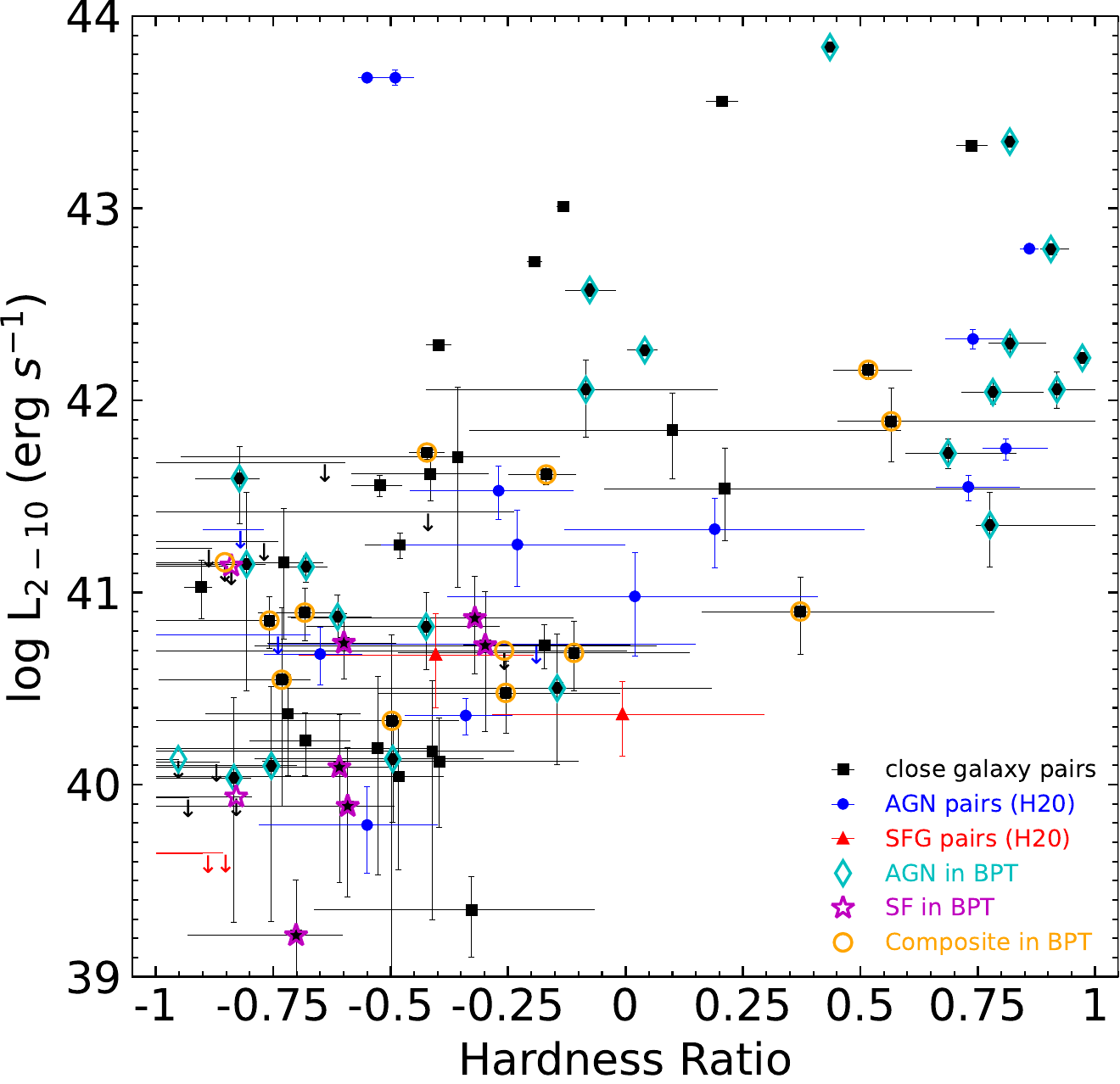}  
\includegraphics[width=0.495\linewidth]{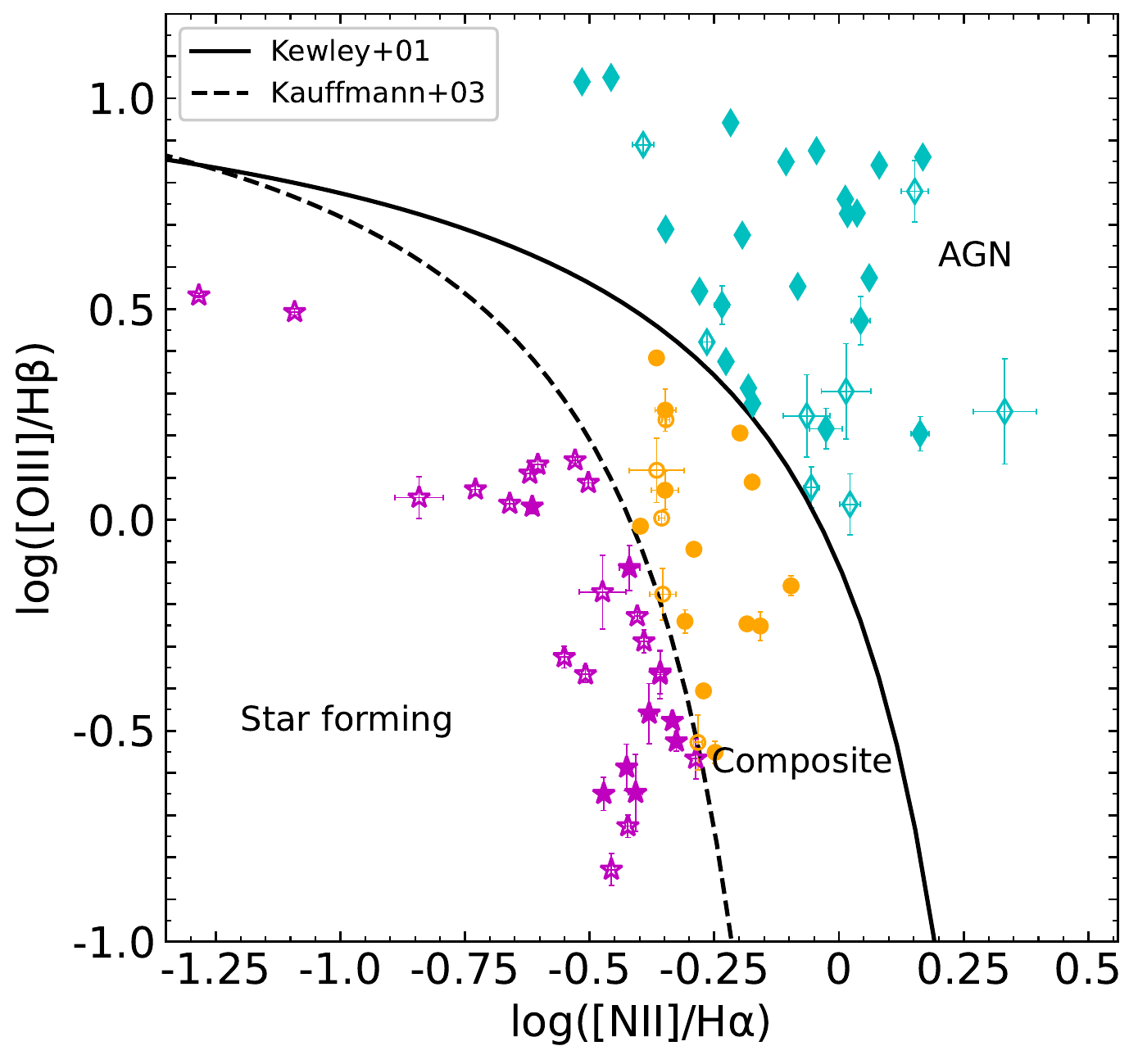}  
\caption{{\it Left}: 2--10 keV luminosity vs. hardness ratio. The black squares, blue circles and red triangles represent X-ray counterparts of close galaxy pairs (current sample), AGN pairs and SFG pairs \citepalias{Hou_2020}, respectively. Those nuclei undetected in the hard band are marked by arrows. 
{\it Right}: Standard BPT diagram for the nuclei which have reliable optical emission line measurements (i.e. with a $S/N > 3$ in each of the four lines). The solid and dashed lines, taken from \citet[]{Kewley_01} and \citet[]{Kaufmann_03}, define the canonical regions occupied by star forming nuclei, composite nuclei and AGNs, which are marked by 
the cyan diamonds, orange circles and magenta stars, respectively (same in the left panel).  
Filled and open symbols represent X-ray detected and non-detected nuclei, respectively.
}
\label{fig:hardness}
\end{figure}

The left panel of Figure \ref{fig:hardness} shows $L_{\rm 2-10}$ against the hardness ratio of the 70 X-ray-detected nuclei in the close galaxy pairs (black squares).
The hardness ratio, which is defined as $HR = (H - S)/(H + S)$, is calculated from the observed photon flux in the $S$ (0.5--2 keV) and $H$ (2--8 keV) bands using a Bayesian approach \citep{Park_2006}. 
For the nuclei that are not detected in the $H$ band, we show the 3$\sigma$ upper limit of $L_{\rm 2-10}$ by arrows in the plot.  
The X-ray counterparts of the AGN pairs (blue circles) and SFG pairs (red triangles) from \citetalias{Hou_2020} are also plotted for comparison (excluding those already included in the new sample).

Sixteen nuclei in the current sample are found to have $L_{\rm 2-10} > 10^{42}{\rm~erg~s^{-1}}$.
These 16 nuclei are probably {\it bona fide} AGNs, but notably only four of them are found in a pair containing another X-ray-detected nucleus (J0907+5203, J1058+3144, J1126+3515 and J1414-0000). 
The majority of close galaxy pairs, however, are found at the bottom left portion with relatively low luminosities ($L_{\rm 2-10} < 10^{41}\rm~erg~s^{-1}$) and a negative $HR$ (i.e., a soft spectrum), a region also occupied by most SFG pairs. 
This may suggest that the X-ray emission of these nuclei are dominated by SF activities (e.g., high-mass X-ray binaries and circumnuclear hot gas heated by supernovae) rather than an AGN. 
However, this does not totally preclude the possibility that some of these sources host an accreting SMBH, either intrinsically weak or heavily obscured by circumnulcear cold gas with a high column density.
In such a case, the observed soft X-rays probably arise further away from the SMBH.

Seventy-five nuclei in the current sample have reliable optical emission line measurements provided by the MPA-JHU SDSS DR7 catalog\footnote{\url{https://wwwmpa.mpa-garching.mpg.de/SDSS/DR7/}}. For these nuclei, we plot a standard BPT diagram \citep[]{BPT_1981}, utilizing the line ratios of \OIII/H$\beta$ and \NII/H$\alpha$ to provide a canonical diagnosis of their nature, namely, SF, AGN or SF/AGN composite, as shown in the right panel of Figure \ref{fig:hardness}. 
Forty-eight of the 75 nuclei can be classified as an AGN or composite, the majority of which are X-ray-detected. 
The remaining 27 nuclei are classified as SF nuclei, but only eight ($\sim30\%$) of them are X-ray-detected, suggesting that the SF activity does not contribute strongly to the observed X-ray emission, at least in this subset of the sample with optical emission line measurements. We note that only two pairs in our sample have both nuclei classified as SF.

We further use SDSS spectroscopic star formation rates (SFR; \citealp{B04}) provided by the MPA-JHU DR7 catalog to estimate the SF-contributed X-ray luminosity. 
We note that the SFR is based primarily on the H$\alpha$ emission line, which might be contaminated in the presence of an AGN, but such an effect should lead to an overestimate of the SF-contributed X-ray luminosity, thus strengthening the following conclusion. 
The information of SFR is available for 137 of the 184 nuclei in the entire sample.
Following \citetalias{Hou_2020}, we adopt the empirical relation of \citet[]{Ranalli_03},
\begin{equation}\label{eq:lxs_sfrs}
L^{{\rm SF}}_{0.5-2} = 4.5 \times 10^{39} \frac{{\rm SFR}}{M_{\odot}~{\rm yr}^{-1}} {\rm erg~s}^{-1}, 
\end{equation}
\begin{equation}\label{eq:lxh_sfrh}
L^{{\rm SF}}_{2-10} = 5.0 \times 10^{39} \frac{{\rm SFR}}{M_{\odot}~{\rm
yr}^{-1}} {\rm erg~s}^{-1},
\end{equation}
which has an rms scatter of 0.27 dex and 0.29 dex in the 0.5--2 keV and 2--10 keV band, respectively.

\begin{figure}[htbp]\centering
\includegraphics[width=0.48\linewidth]{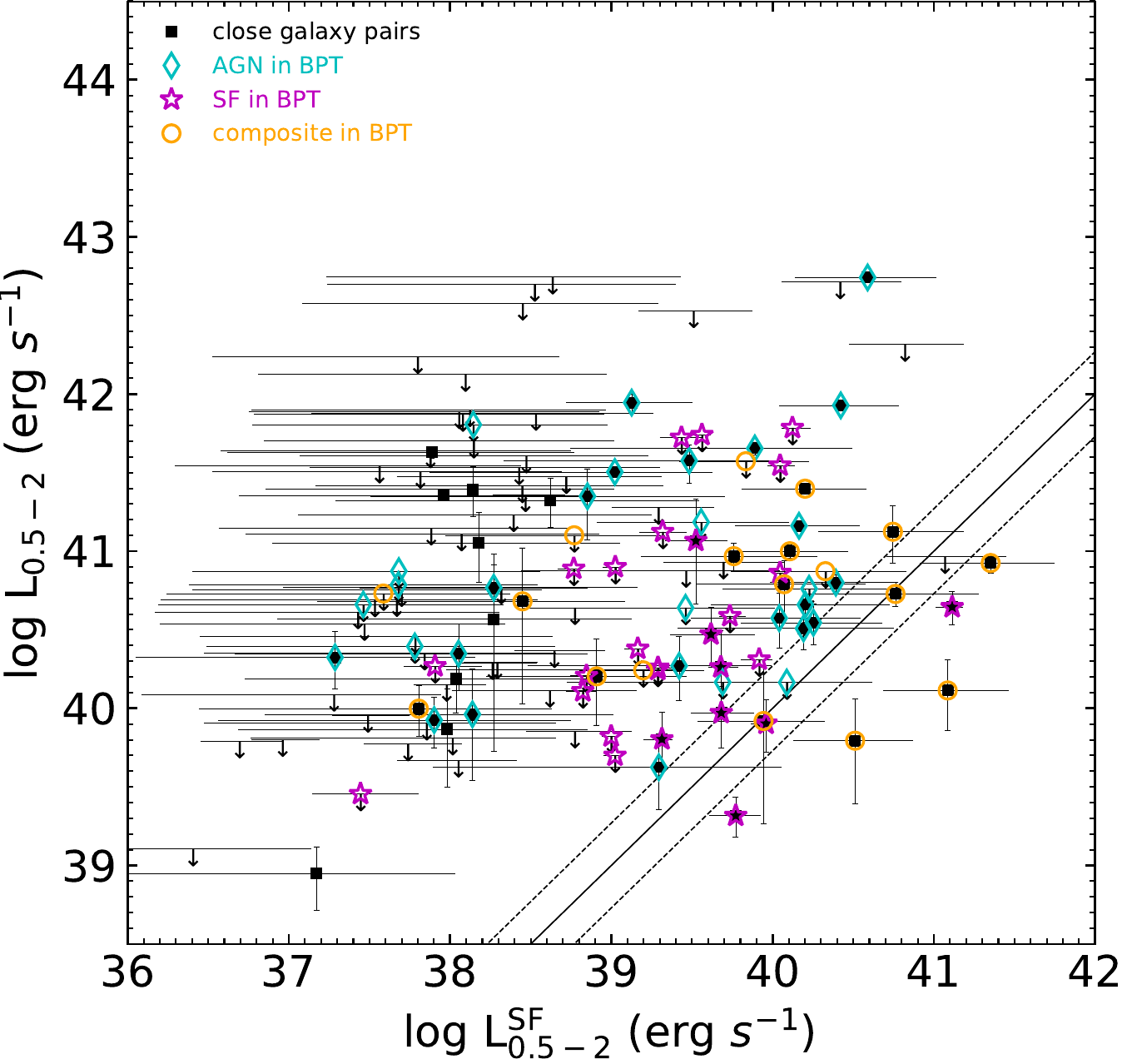}  
\includegraphics[width=0.48\linewidth]{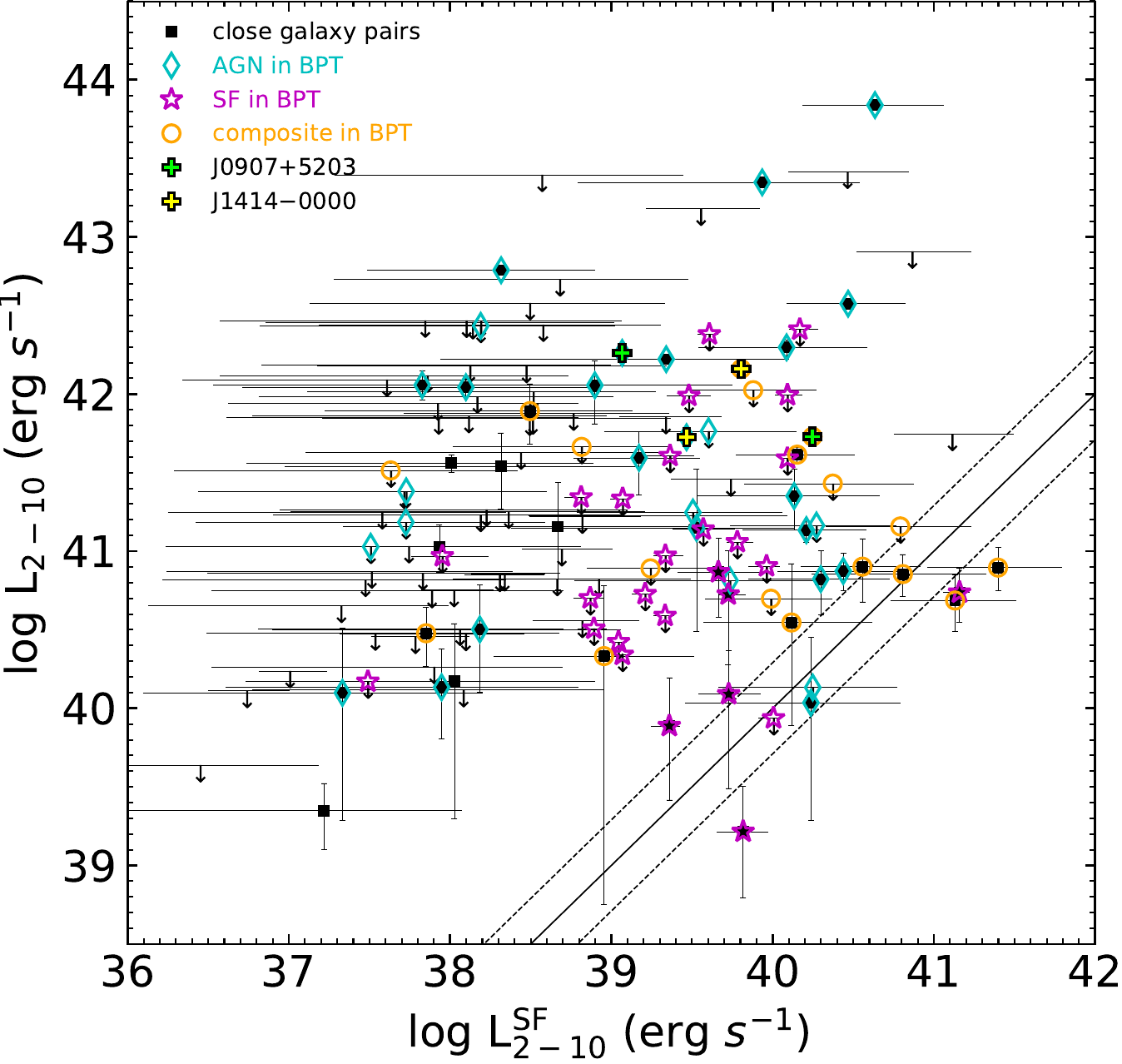}  
\caption{0.5--2 keV ({\it left panel}) and 2--10 keV ({\it right panel}) luminosity versus the predicted luminosity due to star formation activity. 
The black squares 
represent X-ray counterparts of the close galaxy pairs. Those nuclei undetected in a given band are marked by arrows. 
The black solid line indicates a 1:1 relation, with the pair of dashed lines representing the rms scatter (Eqns. \ref{eq:lxs_sfrs} \& \ref{eq:lxh_sfrh}). The cyan diamonds, magenta stars and orange circles denote the optically classified AGNs, SF nuclei and composite nuclei, respectively. 
The two pairs with both nuclei detected above $L_{2-10} = 10^{41}{\rm~erg~s^{-1}}$ are labeled as green and yellow crosses in the right panel.
}
\label{fig:lxsfr}
\end{figure}

Figure \ref{fig:lxsfr} shows the comparison between the measured X-ray luminosity and the empirical X-ray luminosity due to star formation ($L_{0.5-2}^{\rm SF}$, $L_{2-10}^{\rm SF}$) in the two bands. 
The majority of the detected nuclei lie significantly above the predicted SF-contributed luminosity.
This holds in both bands, and more so in the 2--10 keV band.
Still, a few SF nuclei (magenta stars) and a few composite nuclei (orange circles) have their X-ray luminosity consistent with the predicted SF luminosity.
An X-ray AGN is likely absent or heavily obscured in these nuclei.
In the meantime, the majority (but all) of the optically-classified AGN (cyan diamonds) lie significantly above the predicted SF luminosity, indicating that an AGN is indeed powering the observed X-ray emission from these objects.
Overall, Figure~\ref{fig:lxsfr} suggests that $L_{2-10} = 10^{41}{\rm~erg~s^{-1}}$ can be taken as a practical threshold above which a genuine AGN is present and dominates the X-ray emission.  
32 of the 184 nuclei have $L_{2-10}$ above this threshold. Additionally, 83 nuclei have their $3\sigma$ upper limit above this threshold.
We note that only two of the 92 pairs in the current sample (J0907+5203 and J1414-0000, labeled as green and yellow crosses in Figure~\ref{fig:lxsfr}) have both nuclei detected above this threshold.

We also test the SF-contributed X-ray luminosity using different relations provided in \cite{Lehmer2010}, \cite{Mineo2012} and \cite{Fragos2013}. The detailed calculation and figures are presented in Appendix \ref{sec:appendixA}. The overall distributions are very similar to that derived in Figure 5, which help to confirm AGNs dominate the X-ray emission for nuclei with $L_{2-10} = 10^{41}{\rm~erg~s^{-1}}$.

\subsection{Obscured AGNs probed by {\it WISE} color and {\it NuSTAR} spectra}
\label{subsec:obsuredAGN}
\begin{figure}[htbp]\centering
  \includegraphics[width=0.7\linewidth]{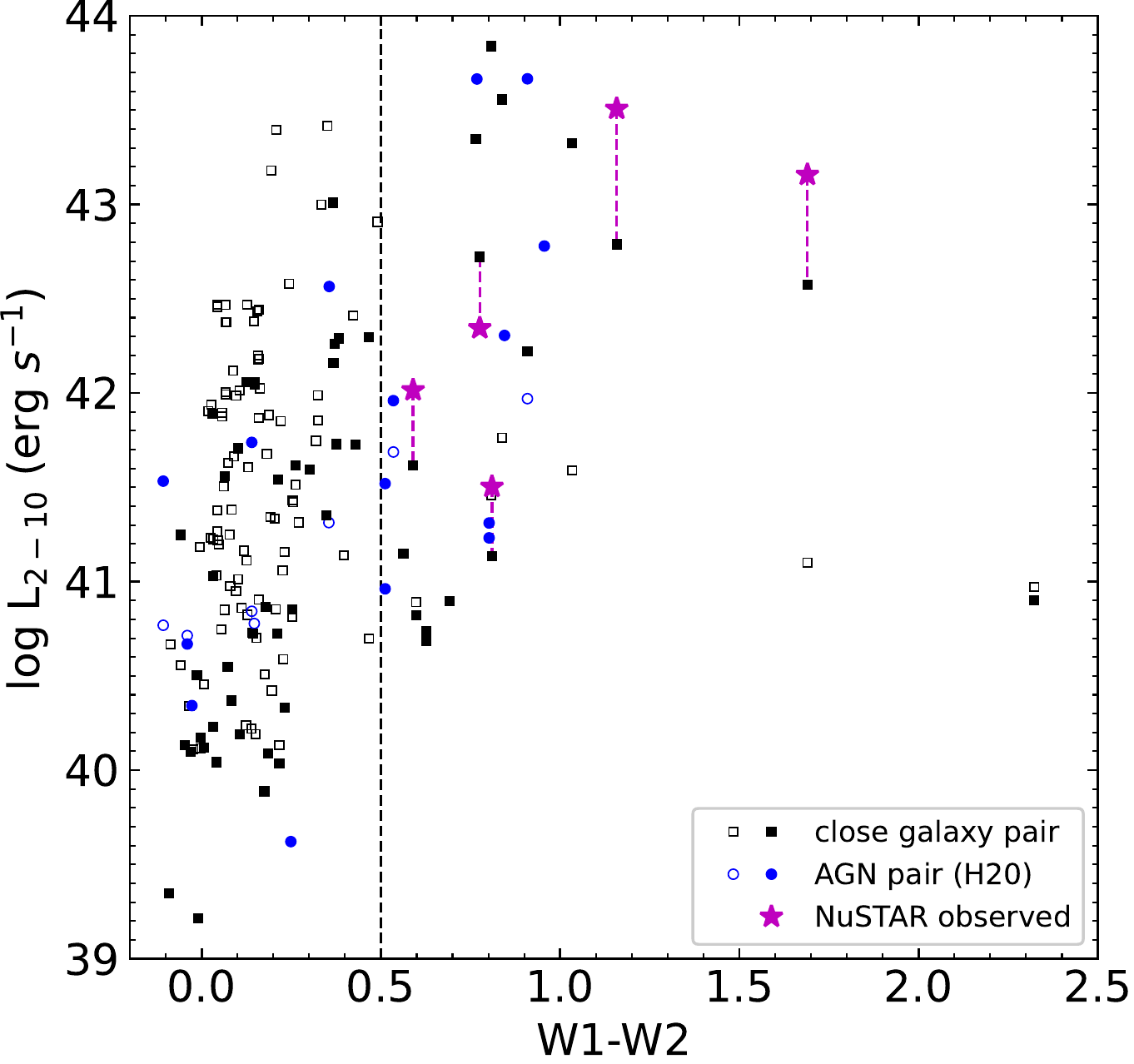}  
\caption{2--10 keV luminosity versus $W1-W2$ color. The close galaxy pairs and AGN pairs of \citetalias{Hou_2020} are shown by black squares and blue circles, respectively. The solid and open symbols represent detections and non-detections. Error bars are neglected for clarity.
The magenta stars mark the 2-10 keV luminosity derived from {\it NuSTAR} spectra, which are available for five pairs.
}
\label{fig:Lh-wise}
\end{figure}

The 2--10 keV luminosity, which is derived by assuming a moderate absorption column density $N_{\rm H} =10^{22}\rm~cm^{-2}$, might be underestimated if the true absorption column density were substantially higher. 
To check the possibility of a buried but intrinsically luminous AGN, we examine the infrared (IR) color of each galaxy pair provided by the Wide-field Infrared Survey Explorer (WISE) survey \citep[]{Wright_10}.
Specifically, we adopt the color of $W1$ (3.4 $\mu$m) $- W2$ (4.6 $\mu$m), which is sensitive to the presence of a luminous AGN \citep[]{Jarrett_11,Stern_12,Satyapal_14}. 
Figure~\ref{fig:Lh-wise} plots $L_{2-10}$ versus $W1 - W2$, for both the close galaxy pairs and AGN pairs. 
Given the relatively large WISE PSF (FWHM $\approx 6\arcsec$), the two nuclei in many of these pairs are unresolved and thus share the same value. Nevertheless, this does not significantly affect our following conclusion, because a luminous AGN, when existed, is expected to dominate the WISE flux. 

Figure~\ref{fig:Lh-wise} shows that most nuclei fall on the blue side of W1--W2 = 0.5, an empirical threshold that separates star-forming galaxies from AGNs \citep[]{Satyapal_14}.
On the other hand, nearly all nuclei with $L_{2-10} > 10^{43}\rm~erg~s^{-1}$ have W1--W2 $> 0.5$, finding good agreement between the X-ray and IR AGN classifications.
A curious exception is the nucleus (J112648.50+351503.2) with the highest $L_{2-10}$ ($1.0\times10^{43}\rm~erg~s^{-1}$), which has W1--W2 $\sim 0.37$, but its high X-ray luminosity warrants an AGN classification, This nucleus is likely accompanied by intense IR starlight of the host galaxy. 
Also remarkable are a handful of nuclei with W1--W2 $> 0.5$ but also with $L_{2-10} \lesssim 10^{43}\rm~erg~s^{-1}$.
Some of these nuclei might host a heavily obscured AGN and have their $L_{2-10}$ significantly underestimated. 
Fortunately, five of these nuclei have an available {\it NuSTAR} spectrum  (Section~\ref{subsec:nustar}).
In four of the five cases (J0841+0102, J1338+4816, J1354+1238 and J1450+0507), 
the 2--10 keV luminosity converted from the best-fit model to the {\it NuSTAR} spectrum is actually 2--7 times higher than the default value of $L_{2-10}$ derived from the {\it Chandra} data (Table \ref{tab:Nuspec}; marked by magenta stars in Figure~\ref{fig:Lh-wise}).
In the remaining case (J1125+5423), the {\it NuSTAR} spectrum-based luminosity is actually 2 times lower, which might reflect intrinsic variability. 
Nevertheless, the absorption column densities inferred from the {\it NuSTAR} spectra are generally moderate, and in all cases lower than $2\times10^{23}\rm~cm^{-2}$ (Table \ref{tab:Nuspec}). This suggests that the $L_{2-10}$ in the other nuclei with W1--W2 $> 0.5$ but without {\it NuSTAR} observations are rather unlikely to have been underestimated by more than a factor of ten. 

\subsection{Mean X-ray Luminosity versus Projected Separation}
\label{subsec:lxrp}

\begin{figure*}[htbp]\centering
\includegraphics[width=0.48\linewidth]{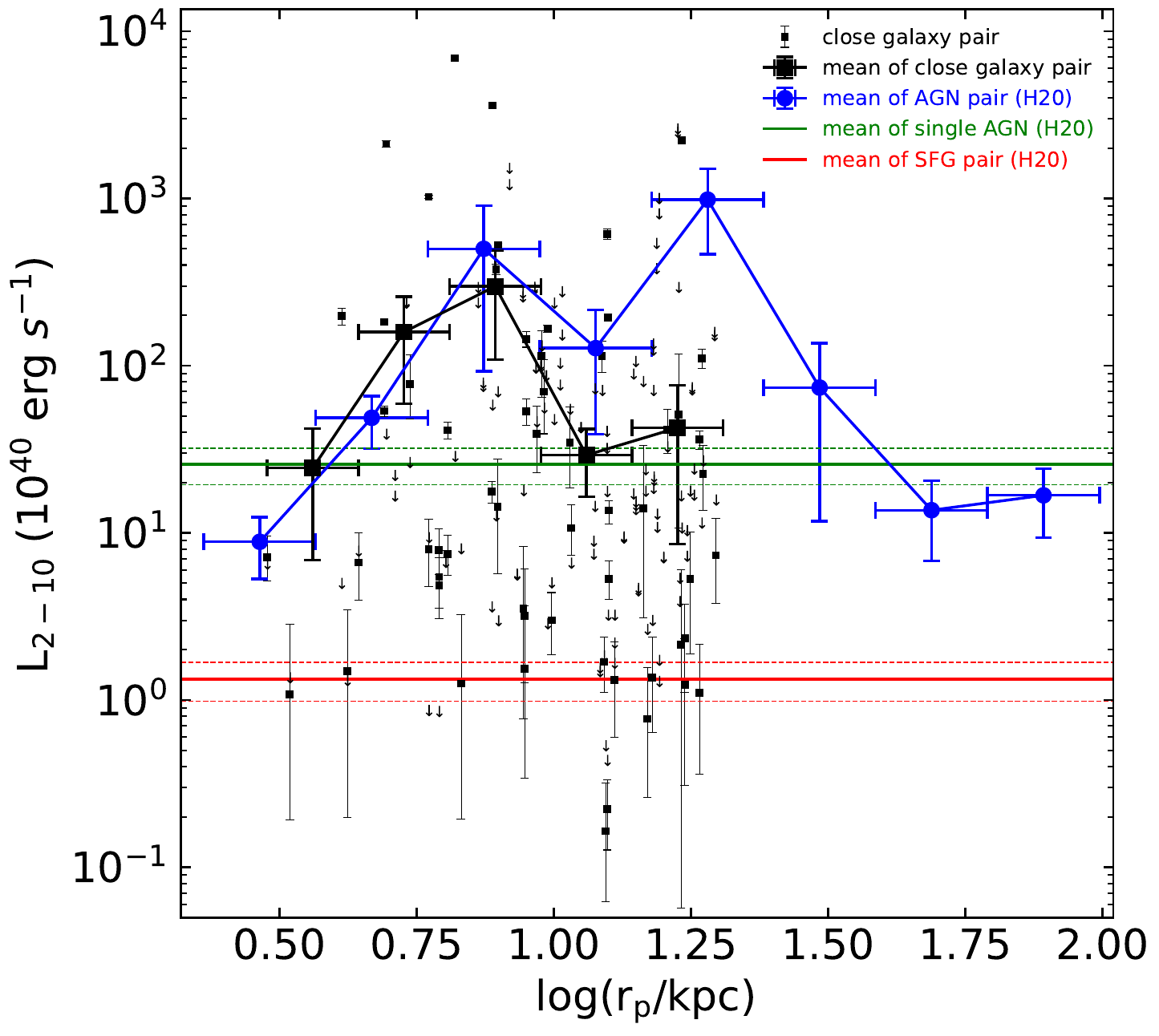} 
\includegraphics[width=0.48\linewidth]{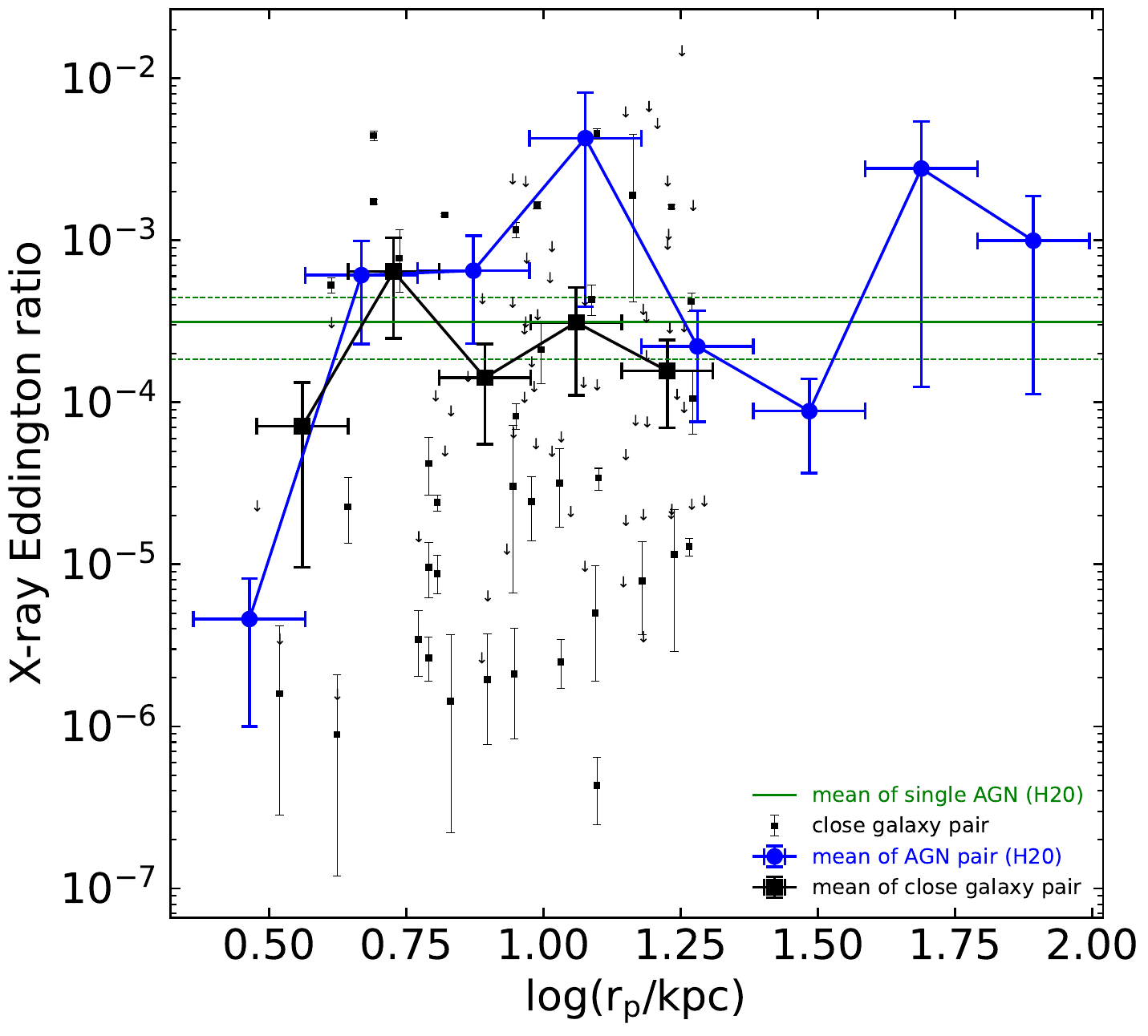} 
\caption{{\it Left}: 2--10 keV luminosity as a function of projected separation. The small black squares and blue circles represent close galaxy pairs and the \citetalias{Hou_2020} AGN pairs, respectively.
The 3$\sigma$ upper limit of undetected nuclei are shown by arrows. For each $r_{\rm p}$ bin, the mean luminosities of close galaxy pairs and the \citetalias{Hou_2020} AGN pairs are represented by the large black squares and blue circles, respectively. The mean value of single AGNs (star-forming galaxy pairs) from \citetalias{Hou_2020} is shown by the green (red) horizontal solid line, with 1$\sigma$ error bars represented by the dashed green (red) lines. 
{\it Right}: Similar to the left panel, but for the X-ray Eddington ratio. The same $r_{\rm p}$ bins as in the left panel are adopted. Only those nuclei with a reliable black hole mass estimate (Table~\ref{tab:info}) are included. 
}
\label{fig:lxrp}
\end{figure*}

The left panel of Figure \ref{fig:lxrp} shows $L_{2-10}$ (or upper limits for non-detected nuclei) as a function of projected separation $r_{\rm p}$ for the close galaxy pairs. 
As mentioned in Section~\ref{sec:intro}, $r_{\rm p}$ is taken as a proxy for the merger phase, with the smallest $r_{\rm p}$ ($\lesssim$ few kpc) indicating the late stage of a merge.
A substantial scatter in $L_{\rm 2-10}$ over nearly 5 orders of magnitude exists in this plot, reflecting a wide range of AGN activity in these galaxies.   
Following \citetalias{Hou_2020}, we bin the data points (including the upper limits) into several intervals of $r_{\rm p}$ and estimate the mean luminosity of each $r_p$ bin using the Astronomy SURVial Analysis (ASURV; \citealp{Feigelson_1985}), a maximum likelihood estimator of the statistical properties of censored data, as is the case here.
We have chosen even bins in logarithmic space covering ${\rm 3.0~kpc} \leq r_{\rm p} \leq {\rm 20~kpc}$ 
 and ensured that each bin contain at least 10 nuclei to minimize random fluctuation.
We note that the main conclusion below is insensitive to the exact choice of bins.
The resultant mean 2--10 keV luminosity of the close galaxy sample is shown by large black squares.
For comparison, the full AGN pair sample of \citetalias{Hou_2020} is shown by blue circles, which covers a wider range of $r_{\rm p}$ up to 100 kpc.
The mean 2--10 keV luminosities of opticall-selected single AGNs and SFG pairs, taken from \citetalias{Hou_2020} and calculated with ASURV, are also plotted for comparison (green and red horizontal lines).

The two outermost bins (10 kpc $ \lesssim r_{\rm p} < 20$ kpc) have a mean $L_{2-10}$ comparable with each other within the statistical uncertainty, which is also comparable to that of optically-selected single AGNs ($2.6[\pm0.6] \times10^{41}\rm~erg~s^{-1}$).
This suggests that galaxy interactions have not generally boosted the AGN activity at such intermediate separations, if the mean X-ray luminosity of single AGNs can be taken as the reference level. 
On the other hand, as noted by \citetalias{Hou_2020} and reiterated here, the AGN pairs at similar $r_{\rm p}$ show a substantially higher mean $L_{2-10}$.
This difference might again be understood as a systematically
higher fraction of luminous AGNs in the \citetalias{Hou_2020} sample, which is pertained to their optical classification.
We note that a handful of nuclei with the lowest $L_{2-10}$ have a value (or upper limit) consistent with the mean of SFG pairs ($1.3[\pm0.3] \times10^{40}\rm~erg~s^{-1}$), indicating that an AGN is intrinsically weak or absent in these nuclei. 

At smaller $r_{\rm p}$, the mean $L_{2-10}$ finds its highest value at the third bin (6.3 kpc $ < r_{\rm p} < 9.0$ kpc), which is about an order of magnitude higher than the mean of the two outer bins as well as the mean of single AGNs.
The mean $L_{2-10}$ of the second bin is also significantly elevated.
This might be understood as a sign of enhanced SMBH accretion due to merger-driven gas inflows. 
However, it is noteworthy that the four nuclei with the highest luminosities ($L_{2-10} \gtrsim 10^{43}\rm~erg~s^{-1}$) were targeted by {\it Chandra} because they were known to be luminous in either hard X-rays or the IR, which potentially introduces a selection effect. 
We find that removing these nuclei from the second and third $r_{\rm p}$ bins results in a mean $L_{2-10}$ much closer to the mean of single AGNs. 
Therefore it remains inclusive whether the upward rising trend between the fourth and third bins is intrinsic. 
More surprisingly, the mean $L_{2-10}$ continues to decrease toward the smallest $r_{\rm p}$.
Nine of the ten nuclei in the innermost bin, in fact, have $L_{2-10}$ (or upper limit) below the mean value of single AGNs. 
Overall, $L_{2-10}- r_{\rm p}$ relation suggests little evidence for merger-induced AGN activity in close galaxy pairs. 

This is reinforced when the absolute X-ray luminosity is replaced by the X-ray Eddington ratio ($L_{2-10}/L_{\rm Edd}$), as shown in the right panel of Figure \ref{fig:lxrp}. 
Here $L_{\rm Edd}$ is the Eddington luminosity, which scales with an estimated black hole mass ($M_{\rm BH}$) based on the stellar velocity dispersion ($\sigma^*$) from the MPA-JHU catalog and the empirical $M_{\rm BH}–\sigma^*$ relation from \citet{Gultekin2009}.
To ensure a reasonable estimate of $M_{\rm BH}$, we have discarded those nuclei with values lower than $10^5\rm~M_\odot$ or higher than 10\% of the host galaxy mass. In total, 91 nuclei have a reliable $M_{\rm BH}$ and appear in the $L_{2-10}/L_{\rm Edd} - r_{\rm p}$ plot. We note that two of the four nuclei with $L_{2-10} > 10^{43}\rm~erg~s^{-1}$ are thus not included.
The mean $L_{2-10}/L_{\rm Edd}$ of the \citetalias{Hou_2020} AGN pairs are plotted for comparison,
as well as the mean $L_{2-10}/L_{\rm Edd}$ of the single AGNs derived in a similar way.
Clearly, the mean $L_{2-10}/L_{\rm Edd}$ of the close galaxy pairs shows no significant enhancement relative to that of single AGNs at any $r_{\rm p}$ bin.

\section{Summary and discussion}\label{sec:discussion}
In this work, we have presented the detection and statistical analyses of X-ray nuclei in a newly compiled sample of 92 close galaxy pairs at low redshift ($\bar{z} \sim 0.07$), based on archival {\it Chandra} observations.
The sample is designed to have projected separations $\lesssim 20$ kpc and thus representative of the intermediate-to-late stage of galaxy mergers. 
Also by design, the sample requires no optical emission line classification of the nuclei, thus it is largely (but not completely) free of selection bias for or against intrinsic AGN activity.
This sample has similar X-ray detection sensitivity (down to a limiting luminosity of $
\sim 10^{40}\rm~erg~s^{-1}$), redshift and host galaxy mass (Figure~\ref{fig:samples}) compared to the close AGN pairs studied by \citetalias{Hou_2020}, but is a factor of $\sim$2 larger in size, helping to relief concern with small number statistics. 
These factors together offer an unprecedented opportunity for probing the connection between galaxy interaction and AGN activity through nuclear X-ray emission, which is generally thought to be a robust diagnostic of AGNs.

Despite the excellent sensitivity achieved, less than half of the 184 nuclei are firmly detected in the X-rays, among which only four have a 2--10 keV unabsorbed luminosity $\gtrsim10^{43}\rm~erg~s^{-1}$, a conventional threshold for luminous AGNs.   
Nevertheless, the majority of the nuclei have an X-ray luminosity (or an upper limit in the case of non-detection) significantly above the empirical luminosity due to star-forming activity  (Figure~\ref{fig:lxsfr}).
This suggests that a weakly accreting SMBH, rather than star formation, is responsible for the observed X-ray emission in most nuclei. 
Optical line ratios, which are available for 75 nuclei, support this view (Figure~\ref{fig:hardness}).

\begin{figure}[htbp]\centering
  \includegraphics[width=0.9\linewidth]{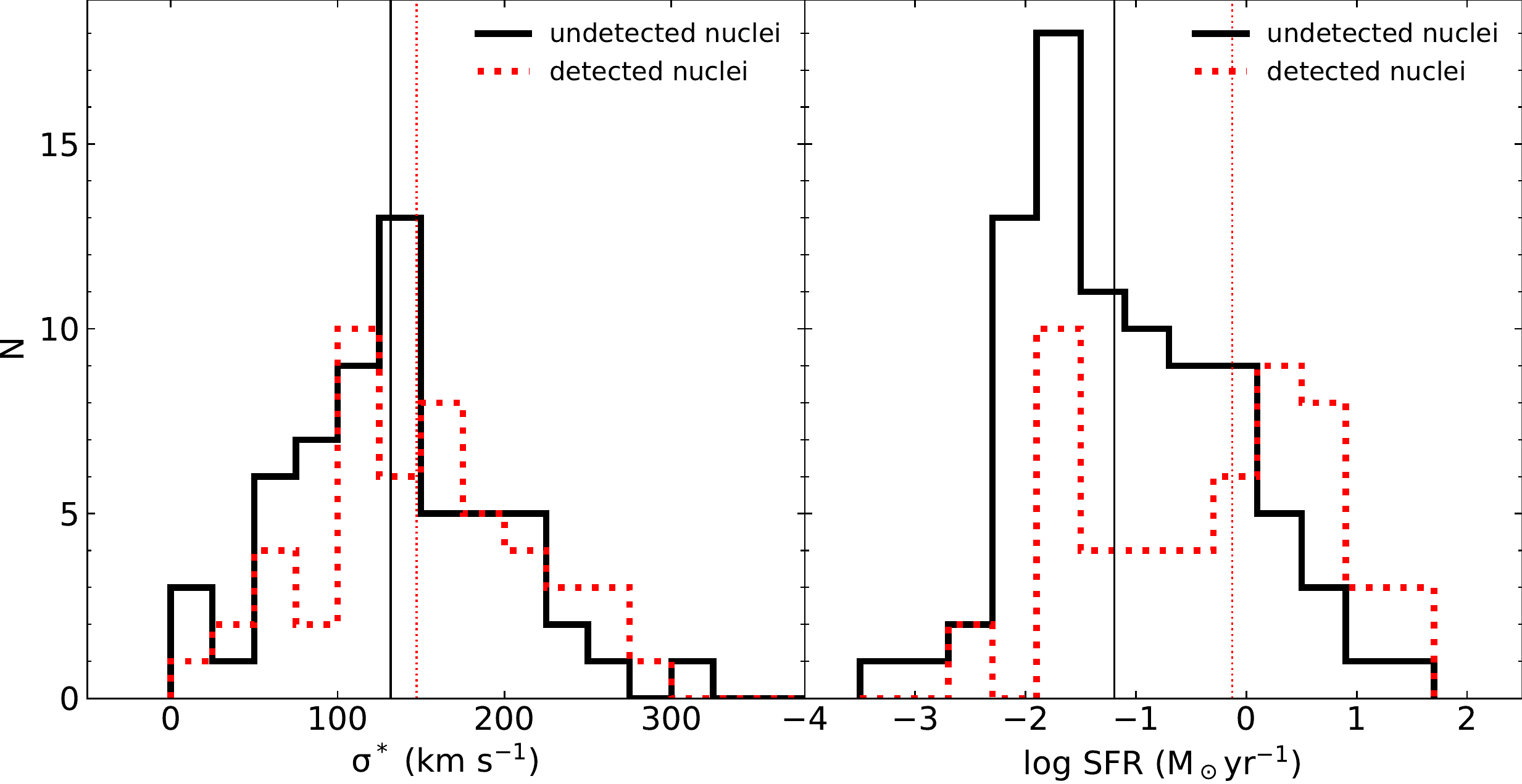}  
\caption{Stellar velocity dispersion (left panel) and star formation rate (right panel) distributions of the X-ray detected (red histogram) and undetected nuclei (black histogram). The vertical lines mark the median values.
}
\label{fig:vdisp-sfr}
\end{figure}

We examine whether the X-ray-detection/non-detections are related to the host galaxy properties. By comparing the distributions of redshift, $r$-band absolute magnitude of the host galaxy, and X-ray detection limit, between the detected and non-detected nuclei, we find that none of these parameters is statistically distinct between the detected and non-detected nuclei.
Our visual examination also does not reveal systematic difference in the global morphology (e.g., more disk-dominated) between the detected and undetected subsets.
In Figure \ref{fig:vdisp-sfr}, we further compare the distributions of stellar velocity dispersion (left panel) and SFR (right panel) between the X-ray detected (red histogram) and undetected nuclei (black histogram). 
The two subsets both show a large scatter in their stellar velocity dispersion, but there is no systematic difference between the two. 
This indicates that the detected nuclei are not preferentially found in galaxies with a more massive SMBH (assuming that $M_{\rm BH}$ is statistically reflected by the stellar velocity dispersion).
On the other hand, a larger fraction of high SFR ($\gtrsim 0.1\rm~M_\odot~yr^{-1}$) is found with the detected subset, although as previously noted the presence of an AGN may cause an overestimate of the SFR.
Neglecting this caveat, such a trend might be taken as evidence that a larger amount of fuel is available in the detected subset for both star formation and SMBH accretion.

We also examine the relation between stellar mass ratio of the pairs and the observed X-ray luminosity. Only about half galaxy pairs (45/92) have reliable stellar mass measurement for both nuclei. Among them, only 27 galaxy pairs have at least one X-ray detected nucleus, which is only a small fraction compared to the whole sample. There is a tentative trend that the more massive galaxy in a pair is more likely to host a more luminous AGN. 

Since essentially all nuclei have a SF-contributed luminosity below $L_{2-10} = 10^{41}\rm~erg~s^{-1}$ (Figure~\ref{fig:lxsfr}), it is practical to adopt this as the threshold, above which a genuine AGN can be identified. 
This allows us to derive 
the fraction of pairs containing at least one X-ray-detected nucleus (the case of only one detected nucleus is sometimes referred to as an ``offset AGN''), which is $\sim33\%$ (Section~\ref{subsec:detectionrate}). 
Raising the threshold to $10^{42}\rm~erg~s^{-1}$ or restricting to dual AGNs (i.e., both nuclei detected) results in a fraction of $17\%^{+5\%}_{-5\%}$ and $2\%^{+2\%}_{-2\%}$, respectively.
These may serve as a useful point of reference for theoretical and numerical studies of AGN triggering in interacting galaxy pairs, in virtue of our sample being largely unbiased to AGN selection. 
Applying different definitions of AGNs (e.g., based on a threshold of bolometric luminosity, X-ray luminosity or Eddington ratio), existing numerical studies, including both idealized galaxy merger simulations (e.g., \citealp{Capelo_15,Capelo_17,Solanes2019}) and cosmological simulations (e.g., \citealp[EAGLE,][]{Rosas_19}; \citealp[ASTRID,][]{Chen_22}), typically predict a dual-AGN fraction of few percent for luminous AGNs or accretion rates close to the Eddington limit. 
This is compatible with the above statistics. 
However, it is noteworthy that current simulations still lack the ability of self-consistently determining the accretion rate and the accretion-induced X-ray luminosity, owing primarily to the lack of resolutions down to the sphere of gravitational influence of the SMBH. 
This is further complicated by the uncertain degree of circumnulear obscuration. 
Hence caution is warranted when comparing the observed and predicted AGN fractions.

\citetalias{Hou_2020} revealed a rather surprising trend of decreasing mean X-ray luminosity with decreasing projected separation in their AGN pairs with $r_{\rm p} \lesssim$ 10 kpc.
This is reproduced in Figure~\ref{fig:lxrp} and further confirmed with the current sample of close galaxy pairs, although one should bear in mind that the innermost bins are driven by a relatively small number of nuclei.
Indeed the mean luminosity of the innermost $r_{\rm p}$ bin is fully consistent with the mean of optically-classified single AGNs.  
The fraction of nuclei with $L_{2-10} > 10^{41}\rm~erg~s^{-1}$, $18\%\pm3\%$ (Table \ref{tab:rate}), is even marginally lower than that of the single AGNs ($24\%\pm5\%$; \citetalias{Hou_2020}).
At face value, this suggests that close galaxy interactions do not {\it effectively} result in boosted AGN activity,
which is contradictive with the general prediction of the aforementioned numerical simulations, in which tidal torques 
become stronger at the smaller separations and thus more effective in driving gas to the vicinity of the SMBH.
Interestingly, a recent study by \citet{Jin_21} based on SDSS/MaNGA integral-field spectroscopic mapping of low-$z$ galaxies, also found no significant excess in the AGN fraction in any merger phase compared to that of isolated galaxies. The galaxy pair sample of \citet{Jin_21} is free of pre-selection of AGN characteristics, which is similar to ours.

Two physical scenarios were proposed by \citetalias{Hou_2020} to explain the behavior revealed in Figure~\ref{fig:lxrp}, which we elaborate here.   
The first is an obscuration effect. In close galaxy pairs, gravitational perturbation can be sufficiently strong to induce gas inflows in one or both galaxies, which in turn result in the accumulation of circumnuclear cold gas that
heavily obscures even the hard X-rays, regardless of the intrinsic AGN luminosity.
Indeed, observational evidence has been gathered for heavily obscured AGN pairs at kpc-separations \citep{Satyapal_17,Pfeifle_19}.
However, obscuration cannot be the sole cause of the low-to-moderate luminosities observed in most nuclei of our sample, in view of the following countering evidences. 
On the one hand, in the five nuclei with a high-quality NuSTAR spectrum, the best-fit foreground absorption column densities (Table~\ref{tab:Nuspec}) are far below that required ($\gtrsim 10^{24}\rm~cm^{-2}$) to completely block X-ray photons below a few keV. 
On the other hand, in a recent attempt of directly detecting circumnuclear cold gas in seven pairs of dual-AGNs based on high-resolution CO observations, \cite{Hou2023} found no evidence for an equivalent hydrogen column density $\gtrsim 10^{24}\rm~cm^{-2}$ in any of the 14 nuclei, which are all included in \citetalias{Hou_2020} (10 included in the current sample).
Nevertheless, it remains interesting to see whether a dense circumnuclear gas exists in the several nuclei with the smallest $r_{\rm p}$ ($\lesssim5$ kpc), which also have the lowest apparent $L_{2-10}$, through higher resolution CO observations and hard X-ray observations.

In the second scenario, most SMBHs in the close galay pairs are currently weakly accreting, which is the result of {\it negative} AGN feedback that have expelled the circumnuclear gas and prevents the SMBH from maintaining a high level of accretion. 
Numerical simulations of idealized galaxy mergers suggest that gas inflows may start as soon as the first pericentric passage of the two galaxies, typically at a physical separation of $\gtrsim10$ kpc, while substantial enhancement of SMBH accretion may not occur until shortly after the second pericentric passage, which lasts for a few tens of Myr \citep{Capelo_15}.
At and after this stage, the separation of the two nuclei remain at no more than 10 kpc, which is consistent with the inner bins in Figure~\ref{fig:lxrp}.
An efficient feedback can explain the moderate column densities inferred for at least a subset of the nuclei.  
The feedback is likely in a kinetic mode mediated by jets and winds \citep{Yuan2014}, given that most nuclei have a low Eddington ratio (Figure~\ref{fig:lxrp}).
Future high-resolution radio and optical spectroscopic observations will be crucial to search for direct evidence of this feedback in the close galaxy pairs.

\section*{Acknowledgements}
M.H. is supported by the National Natural Science Foundation of China (12203001) and National Postdoctoral Program for Innovative Talents of China Postdoctoral Science Foundation (grant BX2021016). 
H.L. and Z.L. acknowledge support by the National Natural Science Foundation of China (12225302). 
S.F. acknowledges support from National Natural Science Foundation of China (No. 12103017) and Natural Science Foundation of Hebei Province (No. A2021205001).
X.L. acknowledges support from NSF grants AST-2108162 and AST-2206499.
The authors wish to thank Drs. Yanmei Chen and Zongnan Li for helpful discussions.

\appendix
\section{Comparison of Star Formation Contribution to X-ray Luminosity}\label{sec:appendixA}
\begin{figure*}[htbp]\centering
\includegraphics[width=0.32\linewidth]{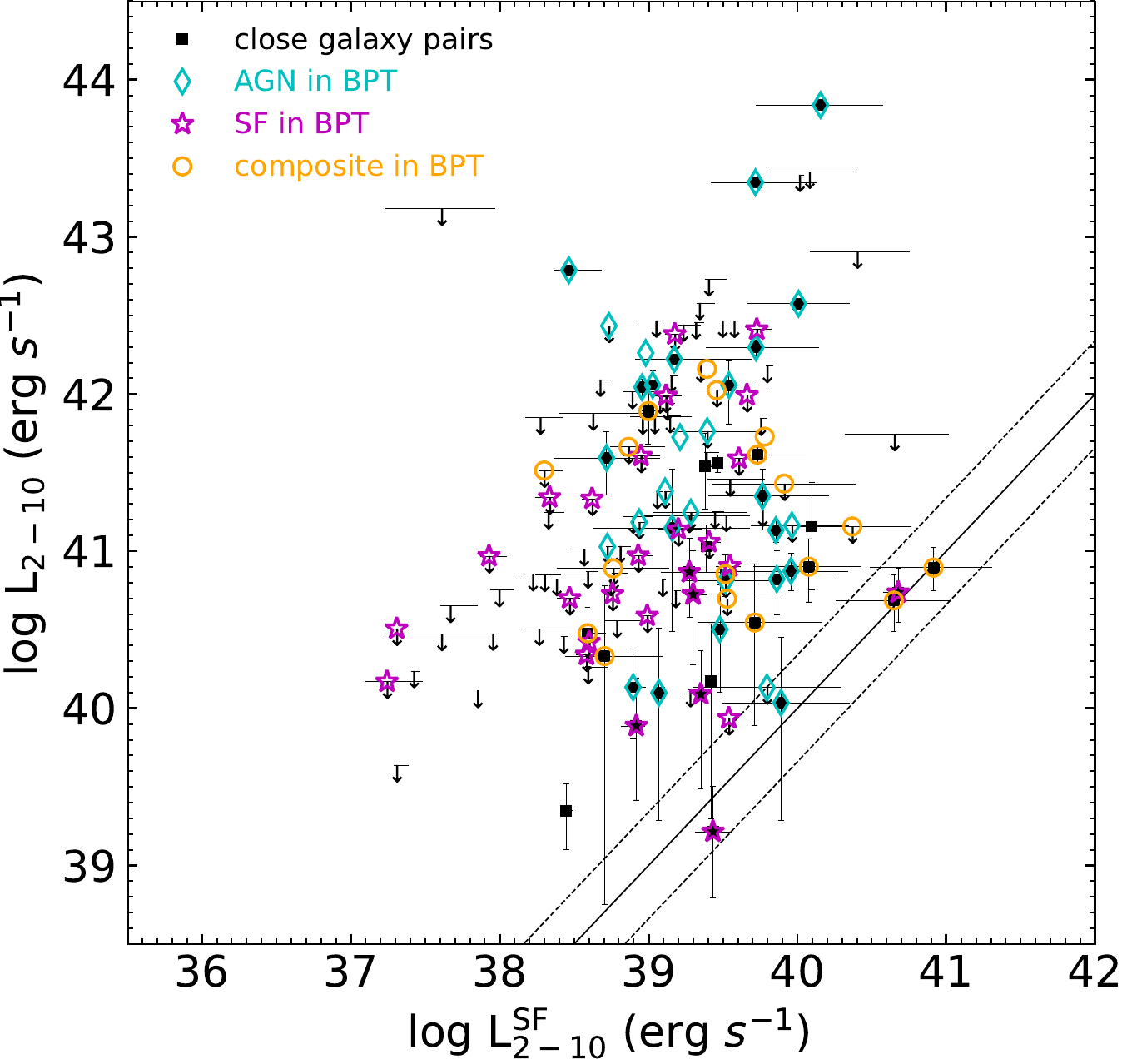}  
\includegraphics[width=0.32\linewidth]{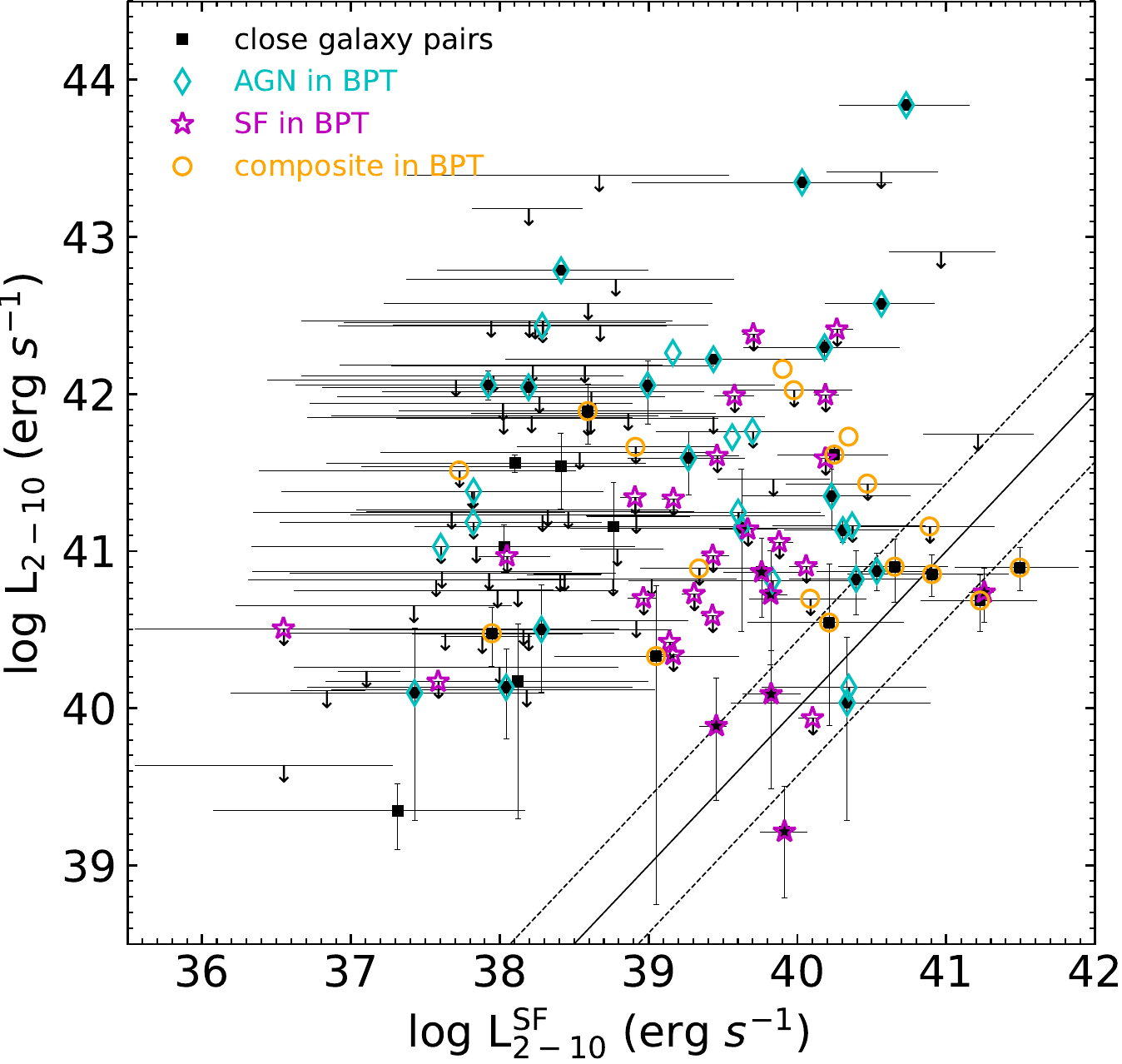}  
\includegraphics[width=0.32\linewidth]{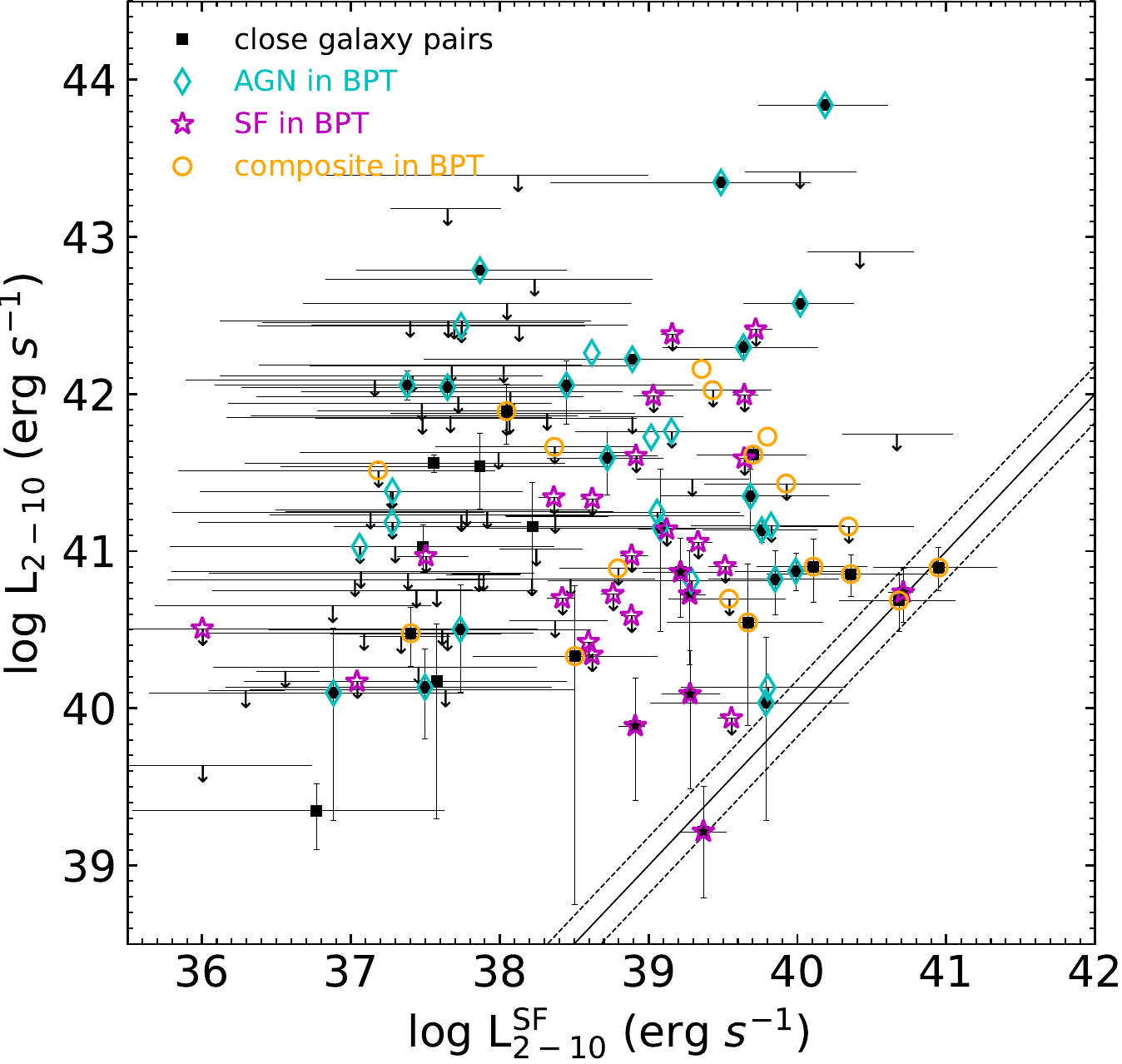}  
\caption{2--10 keV luminosity versus the predicted luminosity due to star formation activity according to the relation in \cite{Lehmer2010} ({\it left panel}), \cite{Mineo2012} ({\it middle panel}), and \cite{Fragos2013} ({\it right panel}), respectively. 
The black squares represent X-ray counterparts of the close galaxy pairs. Those nuclei undetected in a given band are marked by arrows. 
The black solid line indicates a 1:1 relation, with the pair of dashed lines representing the rms scatter. The cyan diamonds, magenta stars and orange circles denote the optically classified AGNs, SF nuclei and composite nuclei, respectively. 
}
\label{fig:lxsfr_compare}
\end{figure*}

\cite{Lehmer2010} calibrated the 2--10 keV X-ray emission from both high- and low-mass X-ray binaries (HMXBs and LMXBs) based on a sample of 17 luminous infrared galaxies and presented an empirical correlation between 2–10 keV luminosity $L^{\rm gal}_{\rm HX}$, SFR, and stellar mass as
\begin{equation}
L^{\rm gal}_{\rm HX} = \alpha M_{*} + \beta {\rm SFR},
\end{equation}
where $\alpha=(9.05\pm0.37)\times10^{28}$ erg s$^{-1}$ $M_{\odot}^{-1}$ and $\beta=(1.62\pm0.22)\times10^{39}$ erg s$^{-1}$ ($M_{\odot}$ yr$^{-1}$)$^{-1}$. 
As estimated based on SDSS images, we adopted a uniform factor of 20\% for the contribution from LMXBs to enclose the stellar mass in the nuclear region ($\sim$ 2\arcsec). Since only half galaxies have stellar mass measurement, the data points are reduced compared to the others (Figure \ref{fig:lxsfr_compare}, left panel).
This relation gives a bit higher 2-10 keV X-ray luminosity because it slightly overestimates the contribution from the pure star-formation related processes. However, nearly all nuclei still lie significantly above the predicted luminosity.

\cite{Mineo2012} considers X-ray emission only from HMXBs with the contamination from LMXBs carefully subtracted based on a sample of 29 nearby star-forming galaxies. But the predicted SF-contributed luminosity is given in 0.5--8 keV as
\begin{equation}
L^{{\rm XRB}}_{{\rm 0.5-8\ keV}} (\rm{erg~s^{-1}}) = 2.61\times10^{39}~{\rm SFR}~(M_\odot\ \rm{yr^{-1}}).
\end{equation}
So we multiply a conversion factor to calculate the 2--10 keV luminosity. As shown in Figure \ref{fig:lxsfr_compare} (middle panel), the distribution is basically the same as that derived in Figure \ref{fig:lxsfr}.

The relations in \cite{Fragos2013} are derived from large scale population synthesis simulations. The X-ray contribution from star-formation related processes are estimated from HMXBs with a dependence on the mean metallicity, given by
\begin{equation}
log(L_{\rm X}/{\rm SFR})=\beta_0 + \beta_1 Z+ \beta_2 Z^2 + \beta_3 Z^3 + \beta_4 Z^4,
\end{equation}
where $\beta_0=42.28\pm 0.02$, $\beta_1=-62.12\pm 1.32$, $\beta_2=569.44\pm 13.71$, $\beta_3=-1833.80\pm 52.14$, $\beta_4=1968.33\pm 66.27$. 
We assume a solar metallicity. The distribution (Figure \ref{fig:lxsfr_compare}, right panel) is also very similar to that derived Figure \ref{fig:lxsfr}, where the majority of the nuclei lie above the predicted SF-contributed luminosity. 

\clearpage

\bibliography{binaryrefs}

\startlongtable
\begin{deluxetable}{cccccccccc}
\tabletypesize{\scriptsize}
\tablecaption{Information of close galaxy pairs with {\it Chandra} observation \label{tab:info}}
\tablewidth{0pt}
\tablehead{
\colhead{Name} & \colhead{R.A.} & \colhead{Dec.} & \colhead{$z$} & \colhead{$r_{\rm p}$} & \colhead{log$M_{\ast}$} &  \colhead{SFR} &  \colhead{log$M_{\rm BH}$} & \colhead{log $L_{\rm X, lim}$} & \colhead{flag}
}
\colnumbers
\startdata
J102700.40+174901.0 & 156.75167 & 17.81694 & 0.0665 & 3.0 & 10.9 & $1.09^{+3.04}_{-0.97}$ & 7.4 & 40.17 & 0 \\
J102700.56+174900.3 & 156.75233 & 17.81675 & 0.0666 & 3.0 & ... & $12.86^{+29.44}_{-9.38}$ & ... & 40.17 & 1 \\
J085837.53+182221.6 & 134.65637 & 18.37267 & 0.0587 & 3.3 & 10.4 & $3.55^{+8.31}_{-2.62}$ & 7.5 & 40.39 & 1 \\
J085837.68+182223.4 & 134.65700 & 18.37317 & 0.0589 & 3.3 & 11.1 & $3.46^{+9.05}_{-2.88}$ & 7.7 & 40.39 & 1 \\
J105842.44+314457.6 & 164.67683 & 31.74933 & 0.0728 & 4.1 & 10.0 & $1.95^{+2.73}_{-1.19}$ & 6.1 & 40.71 & 1 \\
J105842.58+314459.8 & 164.67742 & 31.74994 & 0.0723 & 4.1 & 10.9 & $2.44^{+5.33}_{-1.75}$ & 7.5 & 40.70 & 1 \\
J002208.69+002200.5 & 5.53621 & 0.36681 & 0.0710 & 4.2 & 11.0 & $0.02^{+0.15}_{-0.02}$ & 7.8 & 40.41 & 1 \\
J002208.83+002202.8 & 5.53679 & 0.36744 & 0.0707 & 4.2 & 11.2 & $0.02^{+0.14}_{-0.02}$ & 8.1 & 40.44 & 1 \\
J133031.75-003611.9 & 202.63229 & -0.60331 & 0.0542 & 4.4 & 8.8 & $0.35^{+0.49}_{-0.21}$ & ... & 40.44 & 0 \\
J133032.00-003613.5 & 202.63333 & -0.60375 & 0.0542 & 4.4 & 10.7 & $3.98^{+6.69}_{-2.57}$ & 7.4 & 40.44 & 1 \\
J141447.15-000013.3 & 213.69646 & -0.00369 & 0.0475 & 4.9 & 10.5 & $0.23^{+0.71}_{-0.22}$ & 6.9 & 40.04 & 1 \\
J141447.48-000011.3 & 213.69783 & -0.00314 & 0.0474 & 4.9 & 10.2 & $3.54^{+4.94}_{-2.12}$ & 6.0 & 40.03 & 1 \\
J235654.30-101605.4 & 359.22628 & -10.26817 & 0.0739 & 4.9 & ... & ... & ... & 41.10 & 1 \\
J235654.49-101607.4 & 359.22708 & -10.26875 & 0.0732 & 4.9 & 9.3 & $2.47^{+0.23}_{-0.17}$ & ... & 41.03 & 0 \\
J091931.14+333852.1 & 139.87977 & 33.64782 & 0.0237 & 5.1 & 8.5 & $0.13^{+0.01}_{-0.03}$ & ... & 40.70 & 0 \\
J091930.30+333854.4 & 139.87628 & 33.64845 & 0.0237 & 5.1 & ... & ... & ... & 40.78 & 0 \\
J093529.56+033923.1 & 143.87320 & 3.65644 & 0.0463 & 5.4 & ... & ... & ... & 41.52 & 0 \\
J093529.77+033918.1 & 143.87408 & 3.65505 & 0.0464 & 5.4 & 10.0 & $0.81^{+0.11}_{-0.13}$ & ... & 41.55 & 0 \\
J122814.15+442711.7 & 187.05896 & 44.45325 & 0.0233 & 5.5 & 10.7 & $0.06^{+0.21}_{-0.06}$ & 6.9 & 41.06 & 1 \\
J122815.23+442711.3 & 187.06348 & 44.45314 & 0.0229 & 5.5 & ... & ... & ... & 41.01 & 1 \\
J112648.50+351503.2 & 171.70212 & 35.25089 & 0.0322 & 5.9 & ... & ... & ... & 40.02 & 1 \\
J112648.65+351454.2 & 171.70274 & 35.24839 & 0.0321 & 5.9 & 10.0 & $2.03^{+0.18}_{-0.39}$ & 6.7 & 40.01 & 1 \\
J090025.61+390349.2 & 135.10672 & 39.06369 & 0.0583 & 5.9 & 9.9 & $0.43^{+0.13}_{-0.07}$ & ... & 40.48 & 0 \\
J090025.37+390353.7 & 135.10572 & 39.06492 & 0.0582 & 5.9 & 10.1 & $7.22^{+9.32}_{-4.22}$ & 8.3 & 40.48 & 1 \\
J151806.13+424445.0 & 229.52558 & 42.74585 & 0.0403 & 6.2 & 10.8 & $50.00^{+74.29}_{-31.82}$ & 8.4 & 40.13 & 1 \\
J151806.37+424438.1 & 229.52664 & 42.74387 & 0.0407 & 6.2 & ... & ... & ... & 40.14 & 1 \\
J104518.04+351913.1 & 161.32520 & 35.32032 & 0.0676 & 6.2 & 10.6 & $28.86^{+4.67}_{-6.07}$ & 7.7 & 40.30 & 1 \\
J104518.43+351913.5 & 161.32682 & 35.32041 & 0.0674 & 6.2 & 10.6 & $27.08^{+38.08}_{-16.33}$ & 7.0 & 40.30 & 1 \\
J090332.77+011236.3 & 135.88657 & 1.21009 & 0.0580 & 6.3 & 10.2 & $0.17^{+0.46}_{-0.15}$ & 6.7 & 40.33 & 0 \\
J090332.99+011231.7 & 135.88747 & 1.20881 & 0.0579 & 6.3 & 9.7 & $0.09^{+0.23}_{-0.07}$ & ... & 40.33 & 0 \\
J133817.27+481632.3 & 204.57196 & 48.27564 & 0.0278 & 6.4 & 10.0 & $5.48^{+7.89}_{-3.33}$ & 7.8 & 40.12 & 1 \\
J133817.77+481641.1 & 204.57404 & 48.27808 & 0.0277 & 6.4 & 10.6 & $2.84^{+3.66}_{-1.66}$ & 8.1 & 40.13 & 1 \\
J114753.63+094552.0 & 176.97346 & 9.76444 & 0.0951 & 6.6 & 10.3 & $8.63^{+14.39}_{-5.58}$ & 8.6 & 40.93 & 1 \\
J114753.68+094555.6 & 176.97367 & 9.76544 & 0.0966 & 6.6 & 11.0 & $1.10^{+1.56}_{-0.63}$ & 7.7 & 40.95 & 0 \\
J093634.03+232627.0 & 144.14185 & 23.44083 & 0.0284 & 6.8 & 10.8 & $0.00^{+0.03}_{-0.00}$ & 7.8 & 40.51 & 1 \\
J093633.93+232638.7 & 144.14144 & 23.44411 & 0.0283 & 6.8 & 10.5 & $1.83^{+0.39}_{-0.42}$ & 6.9 & 40.50 & 0 \\
J123257.15+091756.1 & 188.23816 & 9.29892 & 0.1048 & 7.3 & 11.3 & $0.03^{+0.20}_{-0.03}$ & 8.2 & 41.76 & 0 \\
J123257.38+091757.7 & 188.23912 & 9.29939 & 0.1049 & 7.3 & ... & ... & ... & 41.76 & 0 \\
J135853.78+280346.7 & 209.72413 & 28.06300 & 0.0866 & 7.4 & 10.1 & $0.12^{+0.34}_{-0.11}$ & ... & 41.48 & 0 \\
J135853.66+280342.5 & 209.72362 & 28.06182 & 0.0868 & 7.4 & ... & ... & ... & 41.56 & 0 \\
J084113.09+322459.6 & 130.30455 & 32.41657 & 0.0684 & 7.7 & ... & ... & ... & 39.92 & 1 \\
J084112.79+322455.1 & 130.30329 & 32.41533 & 0.0696 & 7.7 & 10.3 & $0.13^{+0.17}_{-0.07}$ & 8.0 & 39.94 & 0 \\
J140737.16+442856.2 & 211.90487 & 44.48229 & 0.1429 & 7.7 & 10.8 & $0.80^{+2.02}_{-0.62}$ & 7.0 & 41.14 & 0 \\
J140737.43+442855.1 & 211.90600 & 44.48200 & 0.1430 & 7.7 & ... & ... & ... & 41.14 & 1 \\
J084135.08+010156.1 & 130.39619 & 1.03228 & 0.1106 & 7.8 & 10.5 & $5.88^{+7.55}_{-3.44}$ & ... & 40.74 & 1 \\
J084134.87+010153.9 & 130.39532 & 1.03165 & 0.1105 & 7.8 & ... & ... & ... & 40.74 & 0 \\
J230010.24-000533.9 & 345.04269 & -0.09276 & 0.1798 & 7.9 & 11.5 & $0.06^{+0.41}_{-0.06}$ & 8.9 & 41.33 & 0 \\
J230010.17-000531.5 & 345.04239 & -0.09211 & 0.1797 & 7.9 & 11.8 & $0.09^{+0.69}_{-0.09}$ & 8.8 & 41.33 & 1 \\
J112536.15+542257.2 & 171.40067 & 54.38264 & 0.0207 & 7.9 & ... & ... & ... & 39.94 & 1 \\
J112535.23+542314.4 & 171.39682 & 54.38741 & 0.0206 & 7.9 & 7.1 & $0.02^{+0.03}_{-0.01}$ & ... & 39.87 & 0 \\
J121247.04+070821.6 & 183.19604 & 7.13933 & 0.1362 & 8.3 & ... & $0.72^{+0.95}_{-0.39}$ & ... & 42.44 & 0 \\
J121246.84+070823.0 & 183.19517 & 7.13975 & 0.1367 & 8.3 & ... & ... & ... & 42.44 & 0 \\
J102229.47+383538.4 & 155.62280 & 38.59401 & 0.0519 & 8.6 & 10.9 & $0.02^{+0.09}_{-0.01}$ & 7.6 & 40.35 & 0 \\
J102229.95+383544.7 & 155.62480 & 38.59577 & 0.0523 & 8.6 & 9.6 & $0.02^{+0.02}_{-0.01}$ & ... & 40.35 & 0 \\
J004343.80+010216.9 & 10.93251 & 1.03805 & 0.1069 & 8.8 & 10.4 & $0.03^{+0.18}_{-0.03}$ & 7.7 & 41.98 & 0 \\
J004344.07+010215.1 & 10.93365 & 1.03754 & 0.1070 & 8.8 & 10.5 & $2.95^{+0.87}_{-0.43}$ & 6.9 & 41.96 & 0 \\
J083817.59+305453.5 & 129.57329 & 30.91486 & 0.0478 & 8.8 & 10.7 & $2.62^{+5.73}_{-1.88}$ & 7.0 & 40.88 & 1 \\
J083817.95+305501.1 & 129.57479 & 30.91697 & 0.0481 & 8.8 & 11.2 & $0.05^{+0.28}_{-0.04}$ & 7.3 & 40.89 & 0 \\
J110713.23+650606.6 & 166.80511 & 65.10192 & 0.0328 & 8.8 & 11.2 & $0.03^{+0.20}_{-0.03}$ & 8.1 & 40.62 & 1 \\
J110713.49+650553.2 & 166.80622 & 65.09819 & 0.0319 & 8.8 & ... & ... & ... & 40.53 & 1 \\
J090714.45+520343.4 & 136.81021 & 52.06206 & 0.0596 & 8.9 & 10.6 & $0.59^{+1.61}_{-0.52}$ & 7.7 & 40.54 & 1 \\
J090714.61+520350.7 & 136.81087 & 52.06408 & 0.0602 & 8.9 & 10.3 & $1.28^{+2.95}_{-0.94}$ & 7.0 & 40.55 & 1 \\
J145309.42+215404.4 & 223.28929 & 21.90123 & 0.1169 & 9.2 & 11.0 & $0.03^{+0.18}_{-0.03}$ & 8.3 & 42.18 & 0 \\
J145309.62+215407.8 & 223.29010 & 21.90220 & 0.1155 & 9.2 & 10.8 & $0.01^{+0.09}_{-0.01}$ & 7.9 & 42.13 & 0 \\
J111828.42-003302.7 & 169.61845 & -0.55077 & 0.1001 & 9.3 & 10.8 & $0.03^{+0.18}_{-0.03}$ & 7.4 & 41.62 & 0 \\
J111828.41-003307.8 & 169.61842 & -0.55216 & 0.1003 & 9.3 & 10.3 & $0.60^{+0.22}_{-0.16}$ & 6.5 & 41.62 & 0 \\
J134736.41+173404.7 & 206.90171 & 17.56797 & 0.0447 & 9.3 & 9.4 & $0.30^{+0.41}_{-0.18}$ & ... & 41.17 & 1 \\
J134737.11+173404.1 & 206.90462 & 17.56781 & 0.0450 & 9.3 & 10.5 & $0.13^{+0.35}_{-0.11}$ & 6.7 & 41.18 & 0 \\
J142445.68+333749.4 & 216.19036 & 33.63039 & 0.0710 & 9.5 & ... & ... & ... & 41.48 & 0 \\
J142445.86+333742.7 & 216.19111 & 33.62855 & 0.0718 & 9.5 & 10.5 & $2.47^{+0.59}_{-0.41}$ & ... & 41.50 & 0 \\
J000249.07+004504.8 & 0.70446 & 0.75133 & 0.0868 & 9.5 & 11.2 & $0.16^{+0.97}_{-0.15}$ & 8.6 & 41.60 & 1 \\
J000249.44+004506.7 & 0.70600 & 0.75186 & 0.0865 & 9.5 & 10.9 & $0.01^{+0.09}_{-0.01}$ & 7.8 & 41.60 & 0 \\
J094543.54+094901.5 & 146.43146 & 9.81709 & 0.1564 & 9.6 & ... & ... & ... & 41.74 & 1 \\
J094543.78+094901.2 & 146.43245 & 9.81700 & 0.1566 & 9.6 & 11.2 & $25.98^{+36.44}_{-14.72}$ & 7.5 & 41.64 & 0 \\
J143106.40+253800.0 & 217.77668 & 25.63335 & 0.0964 & 9.7 & 10.9 & $0.02^{+0.11}_{-0.02}$ & 8.1 & 41.69 & 0 \\
J143106.79+253801.3 & 217.77832 & 25.63370 & 0.0961 & 9.7 & ... & ... & ... & 41.69 & 0 \\
J085953.33+131055.3 & 134.97224 & 13.18205 & 0.0308 & 9.7 & 10.6 & $0.44^{+2.10}_{-0.42}$ & 6.9 & 39.98 & 1 \\
J085952.51+131044.3 & 134.96882 & 13.17900 & 0.0297 & 9.7 & 10.1 & $0.01^{+0.01}_{-0.01}$ & 5.8 & 39.94 & 0 \\
J123515.49+122909.0 & 188.81454 & 12.48585 & 0.0485 & 9.9 & 10.3 & $0.01^{+0.08}_{-0.01}$ & 6.1 & 40.06 & 1 \\
J123516.05+122915.4 & 188.81688 & 12.48763 & 0.0488 & 9.9 & 9.5 & $0.15^{+0.03}_{-0.03}$ & ... & 40.06 & 0 \\
J161758.52+345439.9 & 244.49387 & 34.91109 & 0.1497 & 10.0 & ... & ... & ... & 41.97 & 1 \\
J161758.62+345436.3 & 244.49426 & 34.91007 & 0.1492 & 10.0 & ... & ... & ... & 41.96 & 0 \\
J125253.91-031811.0 & 193.22466 & -3.30309 & 0.0863 & 10.3 & 10.6 & $0.07^{+0.31}_{-0.06}$ & 7.1 & 41.88 & 0 \\
J125254.33-031812.1 & 193.22640 & -3.30338 & 0.0862 & 10.3 & ... & ... & ... & 41.88 & 0 \\
J121514.42+130604.5 & 183.81009 & 13.10126 & 0.1227 & 10.3 & 11.1 & $0.03^{+0.17}_{-0.03}$ & 8.4 & 41.71 & 0 \\
J121514.17+130601.5 & 183.80906 & 13.10043 & 0.1242 & 10.3 & 10.9 & $0.08^{+0.33}_{-0.07}$ & 7.4 & 41.76 & 0 \\
J095749.15+050638.3 & 149.45481 & 5.11066 & 0.1217 & 10.7 & 11.1 & $0.04^{+0.26}_{-0.04}$ & 7.9 & 41.49 & 1 \\
J095748.95+050642.2 & 149.45399 & 5.11174 & 0.1221 & 10.7 & ... & ... & ... & 41.50 & 0 \\
J123637.31+163351.8 & 189.15549 & 16.56441 & 0.0728 & 10.7 & 10.8 & $0.01^{+0.04}_{-0.01}$ & 6.9 & 40.62 & 0 \\
J123637.50+163344.6 & 189.15627 & 16.56239 & 0.0733 & 10.7 & 11.1 & $0.02^{+0.11}_{-0.02}$ & 8.5 & 40.66 & 1 \\
J114608.29-010709.8 & 176.53458 & -1.11940 & 0.1189 & 11.2 & 11.1 & $0.05^{+0.34}_{-0.05}$ & 8.2 & 41.28 & 0 \\
J114608.19-010714.8 & 176.53414 & -1.12078 & 0.1190 & 11.2 & ... & ... & ... & 41.28 & 0 \\
J151110.35+054851.7 & 227.79314 & 5.81437 & 0.0799 & 11.8 & ... & ... & ... & 40.43 & 0 \\
J151109.85+054849.3 & 227.79105 & 5.81370 & 0.0803 & 11.8 & 10.3 & $0.01^{+0.04}_{-0.01}$ & 6.6 & 40.45 & 0 \\
J124545.20+010447.5 & 191.43836 & 1.07987 & 0.1068 & 11.9 & 11.3 & $12.39^{+21.73}_{-8.11}$ & 8.1 & 41.35 & 1 \\
J124545.13+010453.4 & 191.43807 & 1.08153 & 0.1064 & 11.9 & 10.9 & $0.03^{+0.16}_{-0.02}$ & 7.1 & 41.34 & 0 \\
J094130.00+412302.0 & 145.37504 & 41.38390 & 0.0174 & 12.1 & ... & ... & ... & 39.76 & 0 \\
J094132.00+412235.5 & 145.38339 & 41.37656 & 0.0172 & 12.1 & 8.6 & $0.01^{+0.01}_{-0.00}$ & ... & 39.73 & 0 \\
J090134.48+180942.9 & 135.39368 & 18.16195 & 0.0665 & 12.2 & 10.8 & $0.01^{+0.09}_{-0.01}$ & 7.3 & 41.39 & 1 \\
J090135.15+180941.7 & 135.39646 & 18.16159 & 0.0665 & 12.2 & 9.7 & $0.07^{+0.05}_{-0.02}$ & ... & 41.63 & 0 \\
J105622.07+421807.8 & 164.09208 & 42.30219 & 0.0775 & 12.3 & ... & ... & ... & 39.90 & 1 \\
J105622.82+421809.7 & 164.09518 & 42.30267 & 0.0776 & 12.3 & 10.3 & $0.02^{+0.08}_{-0.02}$ & ... & 39.90 & 0 \\
J132924.60+114816.5 & 202.35253 & 11.80459 & 0.0222 & 12.4 & ... & ... & ... & 39.28 & 0 \\
J132924.25+114749.3 & 202.35108 & 11.79703 & 0.0216 & 12.4 & 10.5 & $1.31^{+0.57}_{-0.41}$ & 6.4 & 39.25 & 1 \\
J111136.07+574952.4 & 167.90019 & 57.83131 & 0.0472 & 12.5 & ... & ... & ... & 41.11 & 0 \\
J111134.88+574942.8 & 167.89524 & 57.82866 & 0.0465 & 12.5 & 9.9 & $0.46^{+0.19}_{-0.13}$ & ... & 41.27 & 0 \\
J135429.06+132757.3 & 208.62108 & 13.46592 & 0.0633 & 12.5 & 10.1 & $0.04^{+0.12}_{-0.04}$ & 7.0 & 40.82 & 1 \\
J135429.18+132807.4 & 208.62158 & 13.46872 & 0.0634 & 12.5 & 10.7 & $0.64^{+1.66}_{-0.52}$ & 7.0 & 40.82 & 0 \\
J125725.84+273246.0 & 194.35769 & 27.54613 & 0.0186 & 12.5 & 10.2 & $0.00^{+0.02}_{-0.00}$ & 7.6 & 39.23 & 1 \\
J125723.56+273259.7 & 194.34822 & 27.54993 & 0.0201 & 12.5 & 9.0 & $0.00^{+0.00}_{-0.00}$ & ... & 39.32 & 0 \\
J011448.67-002946.0 & 18.70281 & -0.49612 & 0.0338 & 12.6 & ... & ... & ... & 40.18 & 1 \\
J011449.81-002943.6 & 18.70760 & -0.49542 & 0.0349 & 12.6 & ... & $0.16^{+0.03}_{-0.03}$ & ... & 40.15 & 0 \\
J145051.50+050652.1 & 222.71458 & 5.11448 & 0.0275 & 12.6 & 11.0 & $3.24^{+4.45}_{-1.94}$ & 7.5 & 39.92 & 1 \\
J145050.63+050710.8 & 222.71097 & 5.11968 & 0.0282 & 12.6 & ... & ... & ... & 39.94 & 1 \\
J075311.87+123749.1 & 118.29946 & 12.63031 & 0.0298 & 12.9 & ... & ... & ... & 39.87 & 0 \\
J075313.34+123749.1 & 118.30561 & 12.63031 & 0.0294 & 12.9 & 8.3 & $0.23^{+0.01}_{-0.04}$ & ... & 39.86 & 0 \\
J134844.49+271044.7 & 207.18540 & 27.17911 & 0.0599 & 12.9 & ... & ... & ... & 40.23 & 1 \\
J134844.48+271055.9 & 207.18535 & 27.18220 & 0.0596 & 12.9 & 9.9 & $0.02^{+0.08}_{-0.02}$ & ... & 40.23 & 0 \\
J141958.98+060320.1 & 214.99577 & 6.05560 & 0.0473 & 13.4 & ... & ... & ... & 40.64 & 0 \\
J141959.06+060305.7 & 214.99610 & 6.05161 & 0.0472 & 13.4 & 9.5 & $0.02^{+0.02}_{-0.01}$ & ... & 40.63 & 0 \\
J090005.15+391952.2 & 135.02159 & 39.33120 & 0.0959 & 14.0 & 11.3 & $0.03^{+0.20}_{-0.03}$ & 8.2 & 41.62 & 1 \\
J090005.69+391947.4 & 135.02384 & 39.32988 & 0.0968 & 14.0 & ... & ... & ... & 41.64 & 0 \\
J125315.57-031030.2 & 193.31490 & -3.17507 & 0.0845 & 14.1 & 10.4 & $1.52^{+2.23}_{-0.93}$ & 6.1 & 41.71 & 0 \\
J125315.99-031036.4 & 193.31665 & -3.17680 & 0.0852 & 14.1 & 10.3 & $0.74^{+0.43}_{-0.26}$ & ... & 41.66 & 1 \\
J135225.64+142919.3 & 208.10683 & 14.48869 & 0.0415 & 14.1 & 10.7 & $0.01^{+0.07}_{-0.01}$ & 7.4 & 40.83 & 0 \\
J135226.65+142927.5 & 208.11104 & 14.49097 & 0.0406 & 14.1 & 11.2 & $3.74^{+7.98}_{-2.65}$ & 7.8 & 40.81 & 0 \\
J141807.91+073232.5 & 214.53297 & 7.54237 & 0.0239 & 14.2 & ... & ... & ... & 40.28 & 0 \\
J141805.96+073226.7 & 214.52486 & 7.54077 & 0.0234 & 14.2 & 9.3 & $0.00^{+0.02}_{-0.00}$ & ... & 40.27 & 0 \\
J125400.79+462752.4 & 193.50335 & 46.46458 & 0.0610 & 14.5 & ... & ... & ... & 41.43 & 0 \\
J125359.62+462750.2 & 193.49846 & 46.46397 & 0.0614 & 14.5 & 10.3 & $0.68^{+2.06}_{-0.63}$ & 5.8 & 41.37 & 1 \\
J163026.65+243640.2 & 247.61105 & 24.61118 & 0.0623 & 14.7 & 10.8 & $0.01^{+0.07}_{-0.01}$ & 7.4 & 40.87 & 0 \\
J163026.85+243652.1 & 247.61189 & 24.61449 & 0.0619 & 14.7 & 10.0 & $0.01^{+0.04}_{-0.01}$ & ... & 40.87 & 0 \\
J080133.07+141341.6 & 120.39136 & 14.22618 & 0.0538 & 14.8 & 9.3 & $0.22^{+0.03}_{-0.03}$ & ... & 40.00 & 0 \\
J080133.94+141334.0 & 120.38784 & 14.22821 & 0.0529 & 14.8 & 9.7 & $0.46^{+0.08}_{-0.11}$ & ... & 39.99 & 1 \\
J144804.16+182537.8 & 222.01737 & 18.42718 & 0.0378 & 15.1 & 10.6 & $0.02^{+0.11}_{-0.02}$ & 7.1 & 40.00 & 1 \\
J144804.23+182558.0 & 222.01764 & 18.43277 & 0.0390 & 15.1 & 9.6 & $0.01^{+0.01}_{-0.00}$ & ... & 40.02 & 0 \\
J151031.75+060007.0 & 227.63229 & 6.00195 & 0.0800 & 15.2 & ... & ... & ... & 41.67 & 0 \\
J151031.66+055957.0 & 227.63192 & 5.99919 & 0.0801 & 15.2 & 10.4 & $0.01^{+0.05}_{-0.01}$ & 7.4 & 41.66 & 0 \\
J141115.91+573609.0 & 212.81623 & 57.60258 & 0.1062 & 15.2 & 11.5 & $0.03^{+0.22}_{-0.03}$ & 8.6 & 41.37 & 1 \\
J141115.95+573601.2 & 212.81638 & 57.60041 & 0.1049 & 15.2 & 10.7 & $0.02^{+0.11}_{-0.02}$ & 8.4 & 41.36 & 0 \\
J111627.21+570659.1 & 169.11338 & 57.11651 & 0.0469 & 15.2 & ... & ... & ... & 40.98 & 0 \\
J111625.68+570709.8 & 169.10697 & 57.11950 & 0.0464 & 15.2 & 9.3 & $0.24^{+0.01}_{-0.04}$ & ... & 41.00 & 0 \\
J115532.11+583532.5 & 178.88379 & 58.59246 & 0.1644 & 15.4 & 11.1 & $0.10^{+0.50}_{-0.09}$ & 8.1 & 42.72 & 0 \\
J115532.10+583538.0 & 178.88375 & 58.59397 & 0.1634 & 15.4 & 11.1 & $0.06^{+0.37}_{-0.06}$ & 8.2 & 42.55 & 0 \\
J142553.53+340452.6 & 216.47307 & 34.08129 & 0.0726 & 15.4 & ... & ... & ... & 40.69 & 0 \\
J142553.20+340442.2 & 216.47172 & 34.07840 & 0.0733 & 15.4 & 10.5 & $0.01^{+0.07}_{-0.01}$ & 7.1 & 40.76 & 0 \\
J125917.25-013427.8 & 194.82191 & -1.57440 & 0.1682 & 15.5 & ... & ... & ... & 42.39 & 0 \\
J125917.14-013422.6 & 194.82143 & -1.57297 & 0.1679 & 15.5 & 10.9 & $14.68^{+19.47}_{-8.08}$ & 7.0 & 42.50 & 0 \\
J120429.88+022654.6 & 181.12451 & 2.44849 & 0.0200 & 15.6 & 9.1 & $0.00^{+0.00}_{-0.00}$ & ... & 40.09 & 0 \\
J120432.18+022711.1 & 181.13413 & 2.45310 & 0.0200 & 15.6 & 9.6 & $0.00^{+0.00}_{-0.00}$ & ... & 39.98 & 0 \\
J125922.72+312213.7 & 194.84467 & 31.37050 & 0.0526 & 15.8 & 9.7 & $0.04^{+0.03}_{-0.02}$ & ... & 40.63 & 0 \\
J125922.03+312201.1 & 194.84180 & 31.36698 & 0.0524 & 15.8 & 9.9 & $0.04^{+0.04}_{-0.02}$ & ... & 40.63 & 0 \\
J133525.37+380533.9 & 203.85570 & 38.09276 & 0.0655 & 16.1 & ... & ... & ... & 40.97 & 1 \\
J133525.26+380538.6 & 203.85305 & 38.09515 & 0.0649 & 16.1 & 10.0 & $0.01^{+0.04}_{-0.01}$ & 5.7 & 40.96 & 0 \\
J142442.81-015929.8 & 216.17840 & -1.99163 & 0.1746 & 16.8 & 11.8 & $0.07^{+0.49}_{-0.07}$ & 8.3 & 42.72 & 0 \\
J142442.91-015924.3 & 216.17881 & -1.99011 & 0.1742 & 16.8 & 11.2 & $5.82^{+8.22}_{-3.31}$ & 7.9 & 42.88 & 0 \\
J143541.79+330820.0 & 218.92417 & 33.13891 & 0.1206 & 16.9 & ... & ... & ... & 42.01 & 1 \\
J143542.38+330822.1 & 218.92666 & 33.13947 & 0.1205 & 16.9 & 11.2 & $0.03^{+0.16}_{-0.02}$ & 7.3 & 42.01 & 0 \\
J085405.94+011111.4 & 133.52477 & 1.18650 & 0.0447 & 17.0 & 9.4 & $0.32^{+0.03}_{-0.06}$ & ... & 40.38 & 0 \\
J085405.90+011130.6 & 133.52459 & 1.19186 & 0.0441 & 17.0 & 10.2 & $0.43^{+0.06}_{-0.06}$ & 6.0 & 40.37 & 0 \\
J102108.45+482855.4 & 155.28523 & 48.48206 & 0.0618 & 17.1 & 10.5 & $0.13^{+0.17}_{-0.07}$ & 7.8 & 40.78 & 0 \\
J102109.88+482857.2 & 155.29119 & 48.48256 & 0.0615 & 17.1 & 10.1 & $0.18^{+0.47}_{-0.14}$ & ... & 40.78 & 1 \\
J111519.23+542310.9 & 168.83012 & 54.38636 & 0.0713 & 17.1 & 10.5 & $0.01^{+0.04}_{-0.01}$ & 7.6 & 40.57 & 0 \\
J111519.98+542316.7 & 168.83325 & 54.38797 & 0.0704 & 17.1 & 11.1 & $1.72^{+5.20}_{-1.60}$ & 8.0 & 40.62 & 1 \\
J112402.95+430901.0 & 171.01229 & 43.15028 & 0.0715 & 17.3 & ... & ... & ... & 40.36 & 1 \\
J112401.84+430857.2 & 171.00768 & 43.14922 & 0.0709 & 17.3 & 10.5 & $1.07^{+0.63}_{-0.38}$ & 6.9 & 40.33 & 1 \\
J171255.40+640145.3 & 258.23079 & 64.02934 & 0.0811 & 17.5 & 10.5 & $0.01^{+0.08}_{-0.01}$ & 6.7 & 40.63 & 0 \\
J171255.44+640156.7 & 258.23090 & 64.03252 & 0.0813 & 17.5 & 10.1 & $0.10^{+0.10}_{-0.04}$ & ... & 40.60 & 0 \\
J090215.15+520754.7 & 135.56311 & 52.13189 & 0.1029 & 17.7 & ... & ... & ... & 40.95 & 0 \\
J090215.79+520802.0 & 135.56578 & 52.13393 & 0.1023 & 17.7 & 10.2 & $1.06^{+0.31}_{-0.14}$ & ... & 40.99 & 1 \\
J110418.11+594831.6 & 166.07545 & 59.80882 & 0.1148 & 17.8 & ... & ... & ... & 41.38 & 0 \\
J110419.26+594830.7 & 166.08019 & 59.80861 & 0.1132 & 17.8 & 10.3 & $0.43^{+0.53}_{-0.21}$ & 5.6 & 41.42 & 0 \\
J143454.22+334934.5 & 218.72592 & 33.82625 & 0.0578 & 18.0 & 10.8 & $0.01^{+0.07}_{-0.01}$ & 7.3 & 41.06 & 0 \\
J143454.68+334920.0 & 218.72783 & 33.82222 & 0.0587 & 18.0 & 10.7 & $0.64^{+1.82}_{-0.58}$ & 6.7 & 41.06 & 0 \\
J155207.85+273514.6 & 238.03275 & 27.58740 & 0.0747 & 18.4 & ... & ... & ... & 40.20 & 1 \\
J155207.87+273501.6 & 238.03282 & 27.58380 & 0.0748 & 18.4 & 11.2 & $0.02^{+0.13}_{-0.02}$ & 8.4 & 40.28 & 1 \\
J083902.97+470756.3 & 129.76239 & 47.13233 & 0.0524 & 18.6 & 10.7 & $0.03^{+0.13}_{-0.02}$ & 7.3 & 40.62 & 1 \\
J083902.50+470814.0 & 129.76046 & 47.13722 & 0.0534 & 18.6 & 10.5 & $1.21^{+0.30}_{-0.17}$ & 7.6 & 40.64 & 0 \\
J214622.41+000452.1 & 326.59337 & 0.08114 & 0.0754 & 18.7 & 10.4 & $4.72^{+10.33}_{-3.39}$ & 6.1 & 41.22 & 0 \\
J214623.23+000456.7 & 326.59679 & 0.08242 & 0.0750 & 18.7 & 10.9 & $2.72^{+6.49}_{-2.04}$ & 7.2 & 41.22 & 1 \\
J123042.83+103445.3 & 187.67848 & 10.57926 & 0.1636 & 19.6 & ... & ... & ... & 41.88 & 0 \\
J123043.27+103442.9 & 187.68033 & 10.57860 & 0.1636 & 19.6 & 11.5 & $0.06^{+0.38}_{-0.06}$ & 8.7 & 41.86 & 0 \\
J161111.72+522645.6 & 242.79888 & 52.44607 & 0.0605 & 19.7 & ... & ... & ... & 40.90 & 0 \\
J161113.52+522649.3 & 242.80639 & 52.44709 & 0.0607 & 19.7 & 10.3 & $0.92^{+0.81}_{-0.41}$ & ... & 40.86 & 1 \\
\enddata
\tablecomments{(1) SDSS names with J2000 coordinates given in the form of "hhmmss.ss+ddmmss.s"; (2)-(3) Optical position of the galaxy nucleus; (4) Redshift; (5) Projected physical separation of galaxies in each pair, in units of kpc; (6) Stellar mass, in units of $M_{\odot}$; (7) Star formation rate, in units of $\rm{M}_{\odot}{\rm~{yr}^{-1}}$, given by the MPA-JHU DR7 catalog; (8) Black hole mass estimate inferred from $\sigma_{\ast}$ assuming the $M_{{\rm BH}}$-$\sigma_{\ast}$ relation of \cite{Gultekin2009}, in units of ${\rm M_\odot}$; (9) 0.5-8 keV limiting luminosity for source detection, in units of ${\rm~erg~s^{-1}}$; (10) Flag for X-ray detection, 1 and 0 represent detection and non-detection in X-ray, respectively. 
}
\end{deluxetable}

\clearpage

\startlongtable
\begin{deluxetable}{ccccccccc}
\tabletypesize{\scriptsize}
\tablecaption{X-ray properties of close galaxy pairs  \label{tab:Xray}}
\tablewidth{0pt}
\tablehead{
\colhead{Name} & \colhead{XR.A.} & \colhead{XDec.} & \colhead{Counts} & \colhead{$\rm F_{0.5-2}$} & \colhead{$\rm F_{2-8}$} &  \colhead{log $L_{\rm 0.5-2}$} & \colhead{log $L_{\rm 2-10}$} & \colhead{HR}
}
\colnumbers
\startdata
J102700.56+174900.3 & 156.75230 & 17.81692 & $55.1^{+8.3}_{-8.2}$ & $2.30^{+0.42}_{-0.38}$ & $0.79^{+0.26}_{-0.22}$ & $40.73^{+0.07}_{-0.08}$ & $40.85^{+0.13}_{-0.14}$ & $-0.76^{+0.02}_{-0.24}$ \\
J085837.53+182221.6 & 134.65689 & 18.37266 & $18.7^{+4.9}_{-4.3}$ & $2.55^{+0.67}_{-0.58}$ & $<0.19$ & $40.66^{+0.10}_{-0.11}$ & $<40.13$ & $-0.95^{+0.00}_{-0.05}$ \\
J085837.68+182223.4 & 134.65689 & 18.37266 & $14.2^{+4.4}_{-3.7}$ & $1.79^{+0.57}_{-0.48}$ & $0.15^{+0.25}_{-0.13}$ & $40.51^{+0.12}_{-0.14}$ & $40.04^{+0.42}_{-0.75}$ & $-0.83^{+0.04}_{-0.17}$ \\
J105842.44+314457.6 & 164.67683 & 31.74933 & $2.6^{+2.3}_{-1.6}$ & $0.30^{+0.35}_{-0.23}$ & $<0.46$ & $39.92^{+0.34}_{-0.65}$ & $<40.70$ & $-0.26^{+0.26}_{-0.74}$ \\
J105842.58+314459.8 & 164.67744 & 31.74995 & $96.1^{+10.4}_{-10.3}$ & $1.36^{+0.60}_{-0.48}$ & $18.40^{+2.00}_{-2.10}$ & $40.57^{+0.16}_{-0.19}$ & $42.30^{+0.04}_{-0.05}$ & $0.82^{+0.08}_{-0.05}$ \\
J002208.69+002200.5 & 5.53621 & 0.36681 & $11.9^{+6.0}_{-4.8}$ & $0.58^{+0.29}_{-0.23}$ & $<0.13$ & $40.19^{+0.18}_{-0.22}$ & $<40.12$ & $-0.87^{+0.01}_{-0.13}$ \\
J002208.83+002202.8 & 5.53679 & 0.36744 & $7.8^{+5.2}_{-4.0}$ & $0.28^{+0.22}_{-0.16}$ & $0.14^{+0.19}_{-0.13}$ & $39.87^{+0.26}_{-0.37}$ & $40.17^{+0.37}_{-0.88}$ & $-0.41^{+0.17}_{-0.59}$ \\
J133032.00-003613.5 & 202.63333 & -0.60379 & $17.3^{+4.8}_{-4.1}$ & $2.31^{+0.75}_{-0.63}$ & $1.12^{+0.57}_{-0.45}$ & $40.54^{+0.12}_{-0.14}$ & $40.82^{+0.18}_{-0.23}$ & $-0.42^{+0.16}_{-0.25}$ \\
J141447.15-000013.3 & 213.69646 & -0.00376 & $1188.2^{+36.4}_{-36.0}$ & $27.60^{+1.50}_{-1.50}$ & $40.40^{+1.50}_{-1.50}$ & $41.50^{+0.02}_{-0.02}$ & $42.26^{+0.02}_{-0.02}$ & $0.04^{+0.03}_{-0.04}$ \\
J141447.48-000011.3 & 213.69783 & -0.00321 & $556.3^{+25.0}_{-24.7}$ & $21.70^{+1.40}_{-1.30}$ & $11.90^{+0.80}_{-0.90}$ & $41.40^{+0.03}_{-0.03}$ & $41.73^{+0.03}_{-0.03}$ & $-0.42^{+0.04}_{-0.04}$ \\
J235654.30-101605.4 & 359.22646 & -10.26818 & $526.7^{+24.2}_{-24.0}$ & $24.30^{+3.40}_{-3.40}$ & $188.03^{+9.16}_{-9.06}$ & $41.85^{+0.06}_{-0.07}$ & $43.33^{+0.02}_{-0.02}$ & $0.74^{+0.03}_{-0.03}$ \\
J122814.15+442711.7 & 187.05896 & 44.45325 & $14.3^{+6.3}_{-5.2}$ & $17.80^{+21.20}_{-13.85}$ & $73.90^{+36.53}_{-28.40}$ & $40.68^{+0.34}_{-0.65}$ & $41.89^{+0.17}_{-0.21}$ & $0.57^{+0.43}_{-0.11}$ \\
J122815.23+442711.3 & 187.06348 & 44.45314 & $12.1^{+6.0}_{-4.8}$ & $71.60^{+34.93}_{-27.40}$ & $<25.80$ & $41.27^{+0.17}_{-0.21}$ & $<41.42$ & $-0.42^{+0.18}_{-0.58}$ \\
J112648.50+351503.2 & 171.70213 & 35.25081 & $5273.1^{+76.6}_{-75.9}$ & $469.87^{+9.51}_{-9.41}$ & $500.68^{+10.47}_{-10.37}$ & $42.39^{+0.01}_{-0.01}$ & $43.01^{+0.01}_{-0.01}$ & $-0.13^{+0.01}_{-0.01}$ \\
J112648.65+351454.2 & 171.70263 & 35.24838 & $9.5^{+3.9}_{-3.2}$ & $1.54^{+0.64}_{-0.53}$ & $<0.43$ & $39.90^{+0.15}_{-0.18}$ & $<39.94$ & $-0.83^{+0.03}_{-0.17}$ \\
J090025.37+390353.7 & 135.10567 & 39.06513 & $7.8^{+2.9}_{-2.8}$ & $0.35^{+0.30}_{-0.21}$ & $1.16^{+0.59}_{-0.47}$ & $39.79^{+0.26}_{-0.40}$ & $40.90^{+0.18}_{-0.22}$ & $0.37^{+0.41}_{-0.21}$ \\
J151806.13+424445.0 & 229.52559 & 42.74580 & $76.4^{+9.4}_{-9.3}$ & $10.20^{+1.30}_{-1.35}$ & $2.44^{+0.83}_{-0.69}$ & $40.92^{+0.05}_{-0.06}$ & $40.90^{+0.13}_{-0.14}$ & $-0.68^{+0.07}_{-0.10}$ \\
J151806.37+424438.1 & 229.52648 & 42.74393 & $15.1^{+4.5}_{-3.9}$ & $2.36^{+0.70}_{-0.61}$ & $<0.26$ & $40.30^{+0.11}_{-0.13}$ & $<39.93$ & $-0.93^{+0.00}_{-0.07}$ \\
J104518.04+351913.1 & 161.32538 & 35.32022 & $28.2^{+6.1}_{-5.5}$ & $1.84^{+0.48}_{-0.42}$ & $0.58^{+0.25}_{-0.20}$ & $40.64^{+0.10}_{-0.11}$ & $40.74^{+0.16}_{-0.19}$ & $-0.60^{+0.11}_{-0.16}$ \\
J104518.43+351913.5 & 161.32676 & 35.32023 & $10.6^{+4.2}_{-3.5}$ & $0.55^{+0.31}_{-0.24}$ & $0.52^{+0.24}_{-0.19}$ & $40.11^{+0.19}_{-0.26}$ & $40.69^{+0.16}_{-0.20}$ & $-0.11^{+0.25}_{-0.38}$ \\
J133817.27+481632.3 & 204.57207 & 48.27566 & $92.6^{+10.6}_{-10.5}$ & $16.30^{+2.10}_{-2.00}$ & $4.94^{+1.47}_{-1.25}$ & $40.80^{+0.05}_{-0.06}$ & $40.87^{+0.11}_{-0.13}$ & $-0.61^{+0.07}_{-0.11}$ \\
J133817.77+481641.1 & 204.57415 & 48.27808 & $209.4^{+15.5}_{-15.3}$ & $26.00^{+2.60}_{-2.50}$ & $27.40^{+3.10}_{-3.10}$ & $41.00^{+0.04}_{-0.04}$ & $41.61^{+0.05}_{-0.05}$ & $-0.17^{+0.06}_{-0.08}$ \\
J114753.63+094552.0 & 176.97337 & 9.76444 & $3302.0^{+60.7}_{-60.1}$ & $112.86^{+4.55}_{-4.51}$ & $361.19^{+7.45}_{-7.38}$ & $42.74^{+0.02}_{-0.02}$ & $43.84^{+0.01}_{-0.01}$ & $0.44^{+0.02}_{-0.02}$ \\
J093634.03+232627.0 & 144.14171 & 23.44080 & $7.6^{+3.3}_{-2.7}$ & $5.22^{+2.39}_{-1.93}$ & $0.79^{+1.26}_{-0.67}$ & $40.32^{+0.16}_{-0.20}$ & $40.10^{+0.41}_{-0.81}$ & $-0.75^{+0.05}_{-0.25}$ \\
J084113.09+322459.6 & 130.30458 & 32.41649 & $187.9^{+14.6}_{-14.5}$ & $4.00^{+0.36}_{-0.35}$ & $1.84^{+0.28}_{-0.27}$ & $40.99^{+0.04}_{-0.04}$ & $41.25^{+0.06}_{-0.07}$ & $-0.48^{+0.06}_{-0.07}$ \\
J140737.43+442855.1 & 211.90597 & 44.48199 & $1064.1^{+34.5}_{-34.1}$ & $38.30^{+2.40}_{-2.30}$ & $79.30^{+3.00}_{-3.00}$ & $42.65^{+0.03}_{-0.03}$ & $43.56^{+0.02}_{-0.02}$ & $0.21^{+0.03}_{-0.03}$ \\
J084135.08+010156.1 & 130.39612 & 1.03229 & $366.7^{+20.3}_{-20.1}$ & $12.60^{+1.00}_{-1.00}$ & $14.30^{+1.10}_{-1.00}$ & $41.93^{+0.03}_{-0.04}$ & $42.58^{+0.03}_{-0.03}$ & $-0.08^{+0.06}_{-0.05}$ \\
J230010.17-000531.5 & 345.04272 & -0.09205 & $14.7^{+4.7}_{-4.0}$ & $1.10^{+0.43}_{-0.35}$ & $0.19^{+0.18}_{-0.12}$ & $41.32^{+0.14}_{-0.17}$ & $41.16^{+0.28}_{-0.40}$ & $-0.73^{+0.08}_{-0.21}$ \\
J112536.15+542257.2 & 171.40069 & 54.38269 & $3830.9^{+65.3}_{-64.7}$ & $785.97^{+17.37}_{-17.19}$ & $635.29^{+17.06}_{-16.89}$ & $42.22^{+0.01}_{-0.01}$ & $42.72^{+0.01}_{-0.01}$ & $-0.19^{+0.02}_{-0.02}$ \\
J083817.59+305453.5 & 129.57323 & 30.91485 & $7.8^{+2.9}_{-2.8}$ & $5.26^{+2.17}_{-2.04}$ & $0.77^{+1.05}_{-0.60}$ & $40.79^{+0.15}_{-0.21}$ & $40.55^{+0.37}_{-0.66}$ & $-0.73^{+0.06}_{-0.26}$ \\
J110713.23+650606.6 & 166.80544 & 65.10198 & $4.3^{+2.6}_{-2.0}$ & $1.69^{+1.59}_{-1.05}$ & $1.50^{+1.38}_{-0.90}$ & $39.96^{+0.29}_{-0.42}$ & $40.50^{+0.28}_{-0.40}$ & $-0.15^{+0.33}_{-0.58}$ \\
J110713.49+650553.2 & 166.80633 & 65.09846 & $3.3^{+2.3}_{-1.7}$ & $2.61^{+1.84}_{-1.32}$ & $0.77^{+1.05}_{-0.60}$ & $40.13^{+0.23}_{-0.31}$ & $40.19^{+0.37}_{-0.66}$ & $-0.53^{+0.11}_{-0.47}$ \\
J090714.45+520343.4 & 136.81026 & 52.06206 & $40.7^{+7.1}_{-6.4}$ & $1.01^{+0.54}_{-0.40}$ & $7.36^{+1.39}_{-1.24}$ & $40.27^{+0.19}_{-0.22}$ & $41.73^{+0.08}_{-0.08}$ & $0.69^{+0.15}_{-0.09}$ \\
J090714.61+520350.7 & 136.81087 & 52.06413 & $120.9^{+11.6}_{-11.5}$ & $4.93^{+1.05}_{-0.97}$ & $19.60^{+2.10}_{-2.10}$ & $40.97^{+0.08}_{-0.10}$ & $42.16^{+0.04}_{-0.05}$ & $0.52^{+0.09}_{-0.07}$ \\
J134736.41+173404.7 & 206.90178 & 17.56801 & $67.8^{+8.7}_{-8.6}$ & $86.40^{+11.50}_{-11.50}$ & $9.84^{+4.56}_{-4.12}$ & $41.95^{+0.05}_{-0.06}$ & $41.59^{+0.17}_{-0.24}$ & $-0.82^{+0.04}_{-0.09}$ \\
J000249.07+004504.8 & 0.70433 & 0.75128 & $18.1^{+5.2}_{-6.3}$ & $5.54^{+2.76}_{-2.61}$ & $7.21^{+3.09}_{-3.11}$ & $41.35^{+0.18}_{-0.28}$ & $42.06^{+0.15}_{-0.25}$ & $-0.08^{+0.28}_{-0.34}$ \\
J094543.54+094901.5 & 146.43146 & 9.81709 & $13.5^{+6.4}_{-5.2}$ & $0.74^{+0.78}_{-0.50}$ & $1.27^{+0.71}_{-0.56}$ & $41.02^{+0.31}_{-0.49}$ & $41.84^{+0.19}_{-0.25}$ & $0.10^{+0.49}_{-0.43}$ \\
J085953.33+131055.3 & 134.97212 & 13.18192 & $477.5^{+23.1}_{-22.8}$ & $0.88^{+0.55}_{-0.41}$ & $89.20^{+4.40}_{-4.30}$ & $39.63^{+0.21}_{-0.27}$ & $42.22^{+0.02}_{-0.02}$ & $0.97^{+0.02}_{-0.01}$ \\
J123515.49+122909.0 & 188.81481 & 12.48569 & $31.4^{+9.7}_{-8.4}$ & $0.82^{+0.33}_{-0.28}$ & $0.64^{+0.30}_{-0.24}$ & $40.00^{+0.15}_{-0.18}$ & $40.48^{+0.17}_{-0.21}$ & $-0.25^{+0.24}_{-0.27}$ \\
J161758.52+345439.9 & 244.49387 & 34.91109 & $3.0^{+3.5}_{-2.3}$ & $1.39^{+1.43}_{-0.92}$ & $<0.95$ & $41.25^{+0.31}_{-0.47}$ & $<41.68$ & $-0.64^{+0.04}_{-0.36}$ \\
J095749.15+050638.3 & 149.45481 & 5.11066 & $9.4^{+5.2}_{-4.0}$ & $0.45^{+0.71}_{-0.38}$ & $1.08^{+0.68}_{-0.50}$ & $40.57^{+0.41}_{-0.84}$ & $41.54^{+0.21}_{-0.27}$ & $0.21^{+0.79}_{-0.26}$ \\
J123637.50+163344.6 & 189.15634 & 16.56247 & $163.2^{+13.8}_{-13.7}$ & $15.00^{+1.30}_{-1.30}$ & $0.96^{+0.37}_{-0.30}$ & $41.63^{+0.04}_{-0.04}$ & $41.03^{+0.14}_{-0.16}$ & $-0.90^{+0.02}_{-0.04}$ \\
J124545.20+010447.5 & 191.43838 & 1.08009 & $6.6^{+3.1}_{-2.5}$ & $2.13^{+0.98}_{-0.79}$ & $<0.59$ & $41.12^{+0.16}_{-0.20}$ & $<41.16$ & $-0.85^{+0.01}_{-0.15}$ \\
J090134.48+180942.9 & 135.39369 & 18.16188 & $27.3^{+5.9}_{-5.2}$ & $<1.07$ & $12.60^{+2.90}_{-2.50}$ & $<40.39$ & $42.06^{+0.09}_{-0.10}$ & $0.92^{+0.08}_{-0.01}$ \\
J105622.07+421807.8 & 164.09197 & 42.30219 & $41.1^{+7.4}_{-6.7}$ & $0.56^{+0.11}_{-0.11}$ & $0.14^{+0.05}_{-0.05}$ & $40.25^{+0.08}_{-0.09}$ & $40.23^{+0.15}_{-0.18}$ & $-0.68^{+0.10}_{-0.12}$ \\
J132924.25+114749.3 & 202.35106 & 11.79699 & $16.7^{+4.8}_{-4.2}$ & $0.90^{+0.28}_{-0.24}$ & $0.18^{+0.17}_{-0.11}$ & $39.32^{+0.12}_{-0.13}$ & $39.22^{+0.29}_{-0.42}$ & $-0.70^{+0.10}_{-0.23}$ \\
J135429.06+132757.3 & 208.62108 & 13.46604 & $234.2^{+16.2}_{-16.0}$ & $2.80^{+1.13}_{-0.92}$ & $75.10^{+5.30}_{-5.20}$ & $40.77^{+0.15}_{-0.17}$ & $42.79^{+0.03}_{-0.03}$ & $0.91^{+0.04}_{-0.02}$ \\
J125725.84+273246.0 & 194.35769 & 27.54613 & $21.4^{+7.1}_{-6.4}$ & $0.52^{+0.25}_{-0.22}$ & $0.33^{+0.16}_{-0.14}$ & $38.95^{+0.17}_{-0.23}$ & $39.35^{+0.17}_{-0.24}$ & $-0.33^{+0.26}_{-0.34}$ \\
J011448.67-002946.0 & 18.70286 & -0.49634 & $1097.0^{+35.0}_{-34.6}$ & $163.66^{+6.64}_{-6.57}$ & $86.20^{+4.50}_{-4.40}$ & $41.98^{+0.02}_{-0.02}$ & $42.29^{+0.02}_{-0.02}$ & $-0.40^{+0.03}_{-0.03}$ \\
J145051.50+050652.1 & 222.71453 & 5.11454 & $208.4^{+15.3}_{-15.1}$ & $38.20^{+3.10}_{-3.10}$ & $9.17^{+1.33}_{-1.53}$ & $41.16^{+0.03}_{-0.04}$ & $41.13^{+0.06}_{-0.08}$ & $-0.68^{+0.05}_{-0.05}$ \\
J145050.63+050710.8 & 222.71082 & 5.11957 & $32.9^{+6.4}_{-5.8}$ & $3.81^{+1.08}_{-0.93}$ & $3.41^{+0.96}_{-0.83}$ & $40.18^{+0.11}_{-0.12}$ & $40.73^{+0.11}_{-0.12}$ & $-0.17^{+0.18}_{-0.17}$ \\
J134844.49+271044.7 & 207.18541 & 27.17911 & $10.9^{+4.6}_{-3.9}$ & $0.31^{+0.17}_{-0.13}$ & $0.18^{+0.12}_{-0.10}$ & $39.77^{+0.19}_{-0.24}$ & $40.12^{+0.23}_{-0.34}$ & $-0.40^{+0.30}_{-0.38}$ \\
J090005.15+391952.2 & 135.02133 & 39.33119 & $8.5^{+3.5}_{-2.9}$ & $4.97^{+2.00}_{-1.64}$ & $<0.88$ & $41.39^{+0.15}_{-0.17}$ & $<41.23$ & $-0.89^{+0.01}_{-0.11}$ \\
J125315.99-031036.4 & 193.31665 & -3.17680 & $2.0^{+2.0}_{-1.3}$ & $3.00^{+2.51}_{-1.81}$ & $<0.91$ & $41.07^{+0.26}_{-0.40}$ & $<41.14$ & $-0.84^{+-0.01}_{-0.16}$ \\
J125359.62+462750.2 & 193.49847 & 46.46392 & $15.6^{+3.8}_{-4.3}$ & $19.20^{+4.90}_{-5.40}$ & $1.83^{+2.52}_{-1.43}$ & $41.58^{+0.10}_{-0.14}$ & $41.15^{+0.38}_{-0.66}$ & $-0.81^{+0.04}_{-0.19}$ \\
J080133.94+141334.0 & 120.38814 & 14.22832 & $7.4^{+3.3}_{-2.7}$ & $0.44^{+0.21}_{-0.17}$ & $0.14^{+0.14}_{-0.09}$ & $39.80^{+0.17}_{-0.21}$ & $39.89^{+0.31}_{-0.47}$ & $-0.59^{+0.10}_{-0.41}$ \\
J144804.16+182537.8 & 222.01737 & 18.42721 & $14.0^{+4.4}_{-3.7}$ & $1.16^{+0.47}_{-0.38}$ & $0.48^{+0.36}_{-0.26}$ & $39.93^{+0.15}_{-0.18}$ & $40.13^{+0.24}_{-0.33}$ & $-0.50^{+0.19}_{-0.29}$ \\
J141115.91+573609.0 & 212.81623 & 57.60258 & $20.2^{+5.4}_{-4.8}$ & $1.83^{+1.03}_{-0.80}$ & $<0.76$ & $41.05^{+0.19}_{-0.25}$ & $<41.27$ & $-0.77^{+0.03}_{-0.23}$ \\
J133525.37+380533.9 & 203.85554 & 38.09311 & $62.8^{+11.4}_{-11.2}$ & $9.54^{+2.16}_{-1.90}$ & $4.73^{+1.55}_{-1.31}$ & $41.33^{+0.09}_{-0.10}$ & $41.62^{+0.12}_{-0.14}$ & $-0.42^{+0.12}_{-0.17}$ \\
J143541.79+330820.0 & 218.92417 & 33.13891 & $5.8^{+4.2}_{-3.0}$ & $3.15^{+2.98}_{-1.93}$ & $1.61^{+2.09}_{-1.27}$ & $41.41^{+0.29}_{-0.41}$ & $41.71^{+0.36}_{-0.68}$ & $-0.36^{+0.22}_{-0.59}$ \\
J102109.88+482857.2 & 155.29119 & 48.48256 & $5.2^{+2.9}_{-2.2}$ & $0.81^{+0.59}_{-0.41}$ & $0.28^{+0.51}_{-0.27}$ & $40.20^{+0.24}_{-0.31}$ & $40.33^{+0.45}_{-1.58}$ & $-0.50^{+0.14}_{-0.50}$ \\
J111519.98+542316.7 & 168.83312 & 54.38789 & $1498.0^{+40.9}_{-40.5}$ & $17.30^{+1.70}_{-1.60}$ & $218.16^{+6.22}_{-6.16}$ & $41.65^{+0.04}_{-0.04}$ & $43.35^{+0.01}_{-0.01}$ & $0.82^{+0.02}_{-0.02}$ \\
J112402.95+430901.0 & 171.01226 & 43.15025 & $29.2^{+7.1}_{-6.5}$ & $0.97^{+0.27}_{-0.24}$ & $0.22^{+0.13}_{-0.12}$ & $40.42^{+0.11}_{-0.12}$ & $40.37^{+0.21}_{-0.32}$ & $-0.72^{+0.16}_{-0.18}$ \\
J112401.84+430857.2 & 171.00768 & 43.14922 & $12.0^{+5.2}_{-4.6}$ & $0.35^{+0.17}_{-0.14}$ & $0.12^{+0.11}_{-0.09}$ & $39.97^{+0.17}_{-0.22}$ & $40.09^{+0.28}_{-0.60}$ & $-0.61^{+0.11}_{-0.39}$ \\
J090215.79+520802.0 & 135.56578 & 52.13393 & $5.8^{+3.1}_{-2.4}$ & $0.32^{+0.23}_{-0.16}$ & $0.24^{+0.22}_{-0.15}$ & $40.26^{+0.24}_{-0.31}$ & $40.73^{+0.28}_{-0.45}$ & $-0.30^{+0.37}_{-0.49}$ \\
J155207.85+273514.6 & 238.03275 & 27.58741 & $9.1^{+4.4}_{-3.7}$ & $0.26^{+0.15}_{-0.12}$ & $0.10^{+0.09}_{-0.06}$ & $39.88^{+0.20}_{-0.29}$ & $40.04^{+0.29}_{-0.49}$ & $-0.48^{+0.10}_{-0.52}$ \\
J155207.87+273501.6 & 238.03274 & 27.58389 & $234.2^{+16.3}_{-16.1}$ & $7.68^{+0.64}_{-0.62}$ & $3.14^{+0.40}_{-0.40}$ & $41.36^{+0.03}_{-0.04}$ & $41.56^{+0.05}_{-0.06}$ & $-0.52^{+0.05}_{-0.06}$ \\
J083902.97+470756.3 & 129.76228 & 47.13214 & $70.6^{+8.9}_{-8.8}$ & $1.58^{+0.86}_{-0.67}$ & $20.00^{+2.60}_{-2.60}$ & $40.35^{+0.19}_{-0.24}$ & $42.04^{+0.05}_{-0.06}$ & $0.78^{+0.11}_{-0.07}$ \\
J214623.23+000456.7 & 326.59679 & 0.08242 & $6.3^{+3.3}_{-2.7}$ & $<0.49$ & $1.93^{+0.94}_{-0.76}$ & $<40.16$ & $41.35^{+0.17}_{-0.22}$ & $0.78^{+0.22}_{-0.03}$ \\
J161113.52+522649.3 & 242.80594 & 52.44716 & $10.4^{+3.9}_{-3.2}$ & $1.54^{+0.75}_{-0.59}$ & $0.98^{+0.64}_{-0.48}$ & $40.47^{+0.17}_{-0.21}$ & $40.87^{+0.22}_{-0.29}$ & $-0.32^{+0.21}_{-0.39}$ \\
\enddata
\tablecomments{(1) SDSS names with J2000 coordinates given in the form of "hhmmss.ss+ddmmss.s"; (2)-(3) Centroid position of the X-ray counterpart; (4) Observed net counts in 0.5-8 ($F$) keV bands; (5)-(6) Observed photon flux in 0.5-2 ($S$) and 2-8 ($H$) keV bands, in units of $10^{-6}{\rm~ph~cm^{-2}~s^{-1}}$; (7)-(8) 0.5-2 and 2--10 keV unabsorbed luminosities, in units of ${\rm~erg~s^{-1}}$; (9) Hardness ratio between the 0.5-2 and 2-8 keV bands. 
}
\end{deluxetable}

\clearpage

\begin{deluxetable}{cccccccc}
\tabletypesize{\scriptsize}
\tablecaption{{\it NuSTAR} Spectral Fit Results \label{tab:Nuspec}}
\tablewidth{0pt}
\tablehead{
\colhead{Name} & \colhead{Observation ID} & \colhead{$N_{\rm H}$} & $\Gamma$ & \colhead{$\chi^2/d.o.f$} & \colhead{$F_{3-79}$}  &  \colhead{$L_{2-10,\rm N}$} & \colhead{$L_{2-10,\rm C}$}
}
\colnumbers
\startdata
J0841+0102 & 60401002002 & $0.03^{+7.15}_{-0.03}$ & $0.61^{+0.25}_{-0.13}$ & 37.48/48 & $8.82^{+1.78}_{-2.05}$ & $43.16^{+0.12}_{-0.05}$ & $42.58^{+0.03}_{-0.03}$ \\
J1125+5423 & 60160430002 & $0.78^{+2.50}_{-0.78}$ & $1.59^{+0.13}_{-0.09}$ & 199.88/203 & $8.11^{+0.73}_{-0.67}$ & $42.34^{+0.06}_{-0.04}$ & $42.72^{+0.01}_{-0.01}$ \\
J1338+4816 & 60465005002 & $2.09^{+10.08}_{-2.09}$ & $1.3^{+0.41}_{-0.25}$ & 42.13/38 & $3.17^{+1.03}_{-0.77}$ & $42.02^{+0.19}_{-0.10}$ & $41.61^{+0.05}_{-0.05}$ \\
J1354+1328 & 60160565002 & $17.53^{+4.59}_{-3.99}$ & $1.45^{+0.14}_{-0.13}$ & 166.38/183 & $14.81^{+1.32}_{-1.19}$ & $43.51^{+0.07}_{-0.07}$ & $42.79^{+0.03}_{-0.03}$ \\
J1450+0507 & 60301025002 & $0.02^{+17.78}_{-0.02}$ & $-0.27^{+0.40}_{-0.32}$ & 32.88/35 & $13.08^{+6.10}_{-4.59}$ & $41.50^{+0.18}_{-0.07}$ & $41.13^{+0.06}_{-0.08}$ \\
\enddata
\tablecomments{(1) Source name; (2) {\it NuSTAR} observation ID;  (3) Best-fit column density, in units of $10^{22}\ \rm cm^{-2}$; (4) Best-fit photon index; (5) $\chi^2$ over degree-of-freedom; (6) 3--79 keV unabsorbed flux derived from the best-fit spectral model, in units of $10^{-12}\rm~erg\ s^{-1}\ cm^{-2}$; (7) 2--10 keV intrinsic luminosity derived from the {\it NuSTAR} spectrum; (8) 2--10 keV intrinsic luminosity of the brighter nucleus derived from {\it Chandra} data. 
}
\end{deluxetable}

\begin{deluxetable}{ccccc}
\tabletypesize{\scriptsize}
\tablecaption{Comparison of X-ray Detection Rates \label{tab:rate}}
\tablewidth{0pt}
\tablehead{
\colhead{Sample} & \colhead{Sample Size} & \colhead{Detection Requirement} & \colhead{\# of Detection} & \colhead{Detection Rate}
}
\colnumbers
\startdata
nuclei in close pairs  & 184 & log$L_{\rm 2-10} > 41$ & 32 & $18\%^{+3\%}_{-3\%}$  \\
nuclei with $M_{\rm BH}$  & 91  & $L_{\rm 2-10}/L_{\rm Edd} > 10^{-4}$ &  14 & $15\%^{+4\%}_{-5\%}$  \\
nuclei in H20 AGN pairs (all) & 134 & log$L_{\rm 2-10} > 41$ &  36 & $27\%^{+5\%}_{-5\%}$  \\
nuclei in H20 AGN pairs ($r_{\rm p} \lesssim 20 $ kpc) & 56 & log$L_{\rm 2-10} > 41$ &  22 & $39\%^{+10\%}_{-10\%}$  \\
\hline
close pairs & 92 & at least one detection \& log$L_{\rm 2-10} > 41$ & 30 & $32\%^{+7\%}_{-7\%}$  \\
close pairs & 92 & dual detection \& log$L_{\rm 2-10} > 41$ &  2 & $2\%^{+2\%}_{-2\%}$  \\
close pairs & 92 & at least one detection \& log$L_{\rm 2-10} > 42$ & 16 & $17\%^{+5\%}_{-5\%}$  \\
close pairs ($r_{\rm p} < 10 $ kpc) & 40 & at least one detection \& log$L_{\rm 2-10} > 41$ & 17 & $41\%^{+12\%}_{-13\%}$  \\
close pairs ($r_{\rm p} > 10 $ kpc) & 52 & at least one detection \& log$L_{\rm 2-10} > 41$ & 13 & $25\%^{+8\%}_{-8\%}$  \\
H20 AGN pairs (all) & 67 & at least one detection \& log$L_{\rm 2-10} > 41$ &  32 & $47\%^{+11\%}_{-10\%}$  \\
H20 AGN pairs ($r_{\rm p} \lesssim 20 $ kpc) & 28 & at least one detection \& log$L_{\rm 2-10} > 41$ & 19 & $66\%^{+17\%}_{-18\%}$  \\
close pairs both with $M_{\rm BH}$ & 26 & at least one detection \& $L_{\rm 2-10}/L_{\rm Edd} > 10^{-4}$ & 9  & $33\%^{+13\%}_{-15\%}$  \\
\enddata
\tablecomments{Quoted errors, at 1$\sigma$, take into account the Poisson error associated with both the umerator and denominator.
}
\end{deluxetable}

\end{document}